\documentclass{aa}  
\usepackage{graphicx}
\usepackage{txfonts}
\usepackage{natbib}
\bibpunct{(}{)}{;}{a}{}{,}
\begin{document}
%
 \title{Mass reservoirs surrounding massive infrared dark clouds}
   \subtitle{A view by near-infrared dust extinction}

   \author{J. Kainulainen
          \inst{1}
           J. Alves
          \inst{2}
           H. Beuther
          \inst{1}
           T. Henning
          \inst{1}
          F. Schuller
          \inst{3}
          }

   \offprints{jtkainul@mpia.de}

   \institute{Max-Planck-Institute for Astronomy, K\"onigstuhl 17, 69117
     Heidelberg, Germany \\
              \email{[jtkainul;beuther;henning]@mpia.de}
         \and
             Institute for Astronomy, University of Vienna, T\"urkenschanzstrasse 17, 1180 Vienna, Austria \\
              \email{joao.alves@univie.ac.at}
         \and
             Max-Planck-Institutf\"ur Radioastronomie, Auf dem H\"ugel 69 53121 Bonn, Germany \\
              \email{schuller@mpifr-bonn.mpg.de}
            }

   \date{Received ; accepted }

 
  \abstract
   {Infrared Dark Clouds (IRDCs) harbor progenitors of high-mass stars. Little is known of the parental molecular clouds of the IRDCs.}
   {We demonstrate the feasibility of the near-infrared (NIR) dust extinction mapping in tracing the parental molecular clouds of IRDCs at the distances of $D \approx 2.5 - 8$ kpc. }
   {We derive NIR extinction maps for 10 prominent IRDC complexes using a color-excess mapping technique and NIR data from the UKIDSS/Galactic Plane Survey. We compare the resulting maps to the $^{13}$CO emission line data, to the 8 $\mu$m dust opacity data, and to the millimeter dust emission data. 
We derive distances for the clouds by comparing the observed NIR source densities to the Besan\c{c}on stellar distribution model and compare them to the kinematic distance estimates.}
   {The NIR extinction maps provide a view to the IRDC complexes over the dynamical range of $A_\mathrm{V} \approx 2 - 40$ mag, in spatial resolution of $\sim 30 \arcsec$. The NIR extinction data correlate well with the $^{13}$CO data and probe a similar gas component, but also extend to higher column densities. The NIR data reveal a wealth of extended structures surrounding the dense gas traced by the 8 $\mu$m shadowing features and sub-mm dust emission, showing that the clouds contain typically $\gtrsim 10$ times more mass than traced by those tracers. The IRDC complexes of our sample contain relatively high amount of high-column density material, and their cumulative column density distributions resemble active nearby star-forming clouds like Orion rather than less active clouds like California.}
   {NIR dust extinction data provide a new powerful tool to probe the mass distribution of the parental molecular clouds of IRDCs up to the distances of $D \sim 8$ kpc. This encourages for deeper NIR observations of IRDCs, because the sensitivity and resolution of the data can be directly enhanced with dedicated observations. In addition to mass distributions, the NIR data provide relatively reliable distance estimates.}

   \keywords{ISM: clouds - evolution - ISM: structure - dust, extinction 
               }

  \authorrunning{J. Kainulainen et al.}
  \maketitle


\section{Introduction} 
\label{sec:intro}


The initial conditions of molecular gas from which high-mass stars form remain poorly established. This is, for the most part, due to the violent nature of high-mass star formation which, once started, impacts the cloud structure on a very short timescale. Consequently, the sites where these initial conditions might still be prevailing can only scarcely be found from the nearby high-mass molecular regions such as the Orion, Cepheus, or Vela clouds where the star formation clearly has already taken place (and is still on-going). 


Infrared Dark Clouds (IRDCs) were originally identified as dark structures against the bright mid-infrared background of the Galaxy \citep[][]{per96, ega98}. Due to the details of this technique, the IRDCs do not form a homogenous sample of molecular clouds, but preferably high contrast ($\sim$ high-column density) objects at the distances of $D \lesssim 8$ kpc are detected. The IRDCs are objects in which conditions suitable for high-mass star and star cluster formation are believed to be present or developing \citep[e.g.,][]{rat06, vas09, rat10}. This picture is based on the fact that they contain massive clumps of cold gas ($T\approx 15$ K), up to the masses of young cluster-forming clumps \citep[e.g.,][]{car98, rat06, rat10}. Furthermore, these clumps can contain substructures, i.e. cores, with sizes comparable to pre-stellar \emph{low-mass} cores, but with significantly higher masses (and hence, densities). Some IRDCs have been observed to harbor candidate intermediate- or high-mass protostars \citep[e.g.,][]{pil06, rat10, beu10, hen10}. This demonstrates the capability of IRDCs to form high-mass stars, although to what extent is still under debate \citep[e.g.,][]{kau10letter}. 


The observational technique through which the IRDCs are identified (mid-infrared shadows against the Galactic emission) is sensitive to the material in the column density range $N$(H$_2) \approx 10-100 \times 10^{21}$ cm$^{-2}$ \citep[e.g.,][]{but09, per09, vas09}. Similar sensitivity is reached by mapping thermal dust emission from IRDCs \citep[e.g.,][]{rat06, beu07, rat10, sch09}. Mid-infrared extinction mapping using background sources can reach slightly better sensitivity, but the resolution remains worse due to low density of background sources \citep[108", cf.][]{ryg10}. Observations of CO emission lines can reach relatively low column densities, thereby probing the envelopes of mid-infrared IRDCs \citep[e.g.,][]{her11tan}, but become optically thick, or freeze out \citep[e.g.,][]{her11depletion}, at $N$(H$_2) \gtrsim 5-10 \times 10^{21}$ cm$^{-2}$. As a result, information on the large-scale ($l >$ a few pc) environment of IRDCs, and of the connection of such component to the IRDCs, has remained virtually non-existent. Consequently, some basic questions regarding the parental clouds of IRDCs remain unanswered. To what degree are the IRDCs close to each other in fact components of coherent, large-scale filaments or other structures? Are the molecular clouds harboring IRDCs similar to the nearby high-mass clouds? How much material there is surrounding IRDCs and what is the role of that material for their evolution?


Near-infrared (NIR) observations of stars that shine through molecular clouds provide an alternative technique to probe the low column density, cloud-scale structures surrounding IRDCs. The NIR dust extinction mapping technique is commonly used to estimate gas column densities towards \emph{local} ($r \lesssim 1 $ kpc) molecular clouds \citep[e.g.,][]{lad94, cam02, kai06, kai09b, lom06, lom08, lom10, row09}. Such techniques are based on measuring either the interstellar reddening towards sources that are behind the molecular cloud \citep[color-excess methods, e.g.][]{lad94, lom01, lom09}, or the surface density of such sources \citep[star-count methods, e.g.][]{cam02}. The benefit of such techniques is two-fold. First, the relation between NIR extinction (or reddening) and dust column density depends only relatively weakly on the optical properties of dust grains. Furthermore, it does not depend on dust emissivity index or on dust temperature. Second, the extinction mapping in NIR is sensitive over a considerably different column density range, i.e. $\sim 10^{21} \lesssim N \lesssim 6 \times 10^{22}$ cm$^{-2}$ \citep[e.g.,][]{lom01, kai07, lom08, rom10carlos}, than the mid-infrared shadowing and dust continuum emission observations. Compared to mid-infrared observations, it is also unaffected by local variations in the cloud's surface brightness, thereby being relatively unaffected by signatures of star formation such as nebulae. These properties make near-infrared dust extinction mapping an important complement to mid-infrared extinction and mm dust emission data.


In this paper, we examine the feasibility of using the NIR extinction mapping method to systematically probe the parental molecular clouds of IRDCs that are located at the distances of several kiloparsecs at the Galactic plane where the confusion from diffuse dust is significant. We derive extinction maps for a sample of 10 prominent IRDC complexes using archival NIR data from the UKIDSS/Galactic Plane Survey \citep[][hereafter GPS]{luc08} and compare them to $^{13}$CO emission line data from the Boston University-FCRAO Galactic Ring Survey \citep[][hereafter GRS]{jac06}, mid-infrared opacity data from the recently published IRDC catalogue by \citet{per09}, and to sub-millimeter dust continuum data from the ATLASGAL survey \citep{sch09}. This combination of data will be used to examine the mass distribution surrounding the IRDCs and the fraction of dense cold gas in their parental molecular clouds. We will also use the NIR data to derive distances for our IRDC sample by determining the density of sources on the foreground to the IRDCs and comparing that to the predicted source density from the Besan\c{c}on Galactic stellar distribution model \citep{rob03}. 

In Section \S\ref{sec:methods} we introduce the data used in the paper and the extinction mapping technique. In Section \ref{sec:distances} we present the distance determination method and apply it to the sample. Section \S\ref{sec:results} presents our results, which are then discussed in \S\ref{sec:discussion}. In Section \S\ref{sec:conclusions} we give our conclusions. In addition, Appendix \ref{app:figures} gives the extinction maps for the sample, together with the $^{13}$CO, 8 $\mu$m opacity, and sub-mm dust emission data.

\section{Methods and data}                     
\label{sec:methods}

\subsection{Sample of IRDC complexes}
\label{subsec:sample}

We selected the objects for this study by visually inspecting the Spitzer/GLIMPSE survey images and choosing from them a sample of spatially extended complexes with high contrast against the background. The sample selection was restricted, on the one hand, by the coverage of the UKIDSS/GPS that was used as NIR data in the study, and on the other hand, by the coverage of the GLIMPSE survey and the contrast of complexes in it. These restrictions resulted in selection of complexes between $l \approx 0-40\deg$ and $|b| < 1\deg$. One of the main objectives of our study is to test the practical limits of the NIR extinction mapping technique in mapping distant clouds. Therefore, we selected complexes located both relatively near (e.g., 35.49-00.31, $D_\mathrm{grs} = 3.0$ kpc) and far (e.g., 32.09+00.09, $D_\mathrm{grs}=7.1$ kpc) based on kinematic distance estimation from the GRS data. As a result, the sample does not form a well-defined selection, although the chosen clouds are likely among the largest (most massive) IRDC complexes at the chosen spatial range. Nine out of ten sample clouds have counterparts in the 8 $\mu$m opacity catalogue, and eight in the BU/GRS survey. All complexes are included in the area surveyed by ATLASGAL. The chosen complexes are listed in Table \ref{tab:clouds} and designated according to the notation of the GRS survey \citep{rom10}.

\subsection{Near-infrared extinction mapping}
\label{subsec:nicer}


We employed the color-excess mapping
technique {\sf nicer} \citep{lom01} to derive dust 
extinction maps for the sample of IRDC complexes. The method was used
in conjunction with $JHK_S$ band photometric data from the UKIDSS/GPS \citep{luc08}. In particular, data from the Data Release 7 Plus were used. The data reach approximately the limiting magnitudes of $K=19$ mag, $H=19.1$ mag, and $J=20$ mag. The UKIDSS project is defined in \citet{law07}. UKIDSS uses the UKIRT Wide Field Camera \citep[WFCAM,][]{cas07}. The photometric system is described in \citet{hew06}, and the calibration is described in \citet{hod09}. The science archive is described in \citet{ham08}.

In the  basic implementation of the {\sf nicer} extinction mapping technique the NIR colors of stars, shining through molecular clouds, are compared to the colors of stars in a nearby control field that is free from
extinction. This comparison yields measurements of a NIR reddening
towards the stars in the molecular cloud region, from which extinction can be estimated by adopting a wavelength dependency for the reddening, i.e. a reddening law. The extinction values derived towards each source ($A_\mathrm{V}^*$) are then used to compute a spatially smoothed map of dust extinction through the cloud ($A_\mathrm{V}$). This basic approach is straightforward to apply for \emph{nearby}  molecular clouds ($D \lesssim 500$ pc), in which case almost all sources that are detected are located behind the intervening dust cloud. However, in the case of IRDCs which are known to be at the distances of several kiloparsecs, a significant fraction of the detected stars are located between the cloud and the observer, i.e. on the foreground to the cloud. Such stars bias the extinction determination at low column densities and totally hamper it at high column densities \citep[e.g.,][]{lom05}. Optimally, the foreground sources should all be eliminated from the sources used for extinction mapping. In practice, excluding them uniformly from everywhere is not possible due to the spread in the intrinsic colors of stars which mimics dust extinction. In the following, we introduce the practical implementation of the \textsf{nicer} technique used in this paper. For the description of the basics of the method itself, we refer to \citet{lom01} \citep[see also][who presents an analysis of the effect of foreground stars for the technique]{lom05}. 

\begin{figure}
   \centering
\includegraphics[bb = 50 290 290 480, clip=true, angle=270, width=\columnwidth]{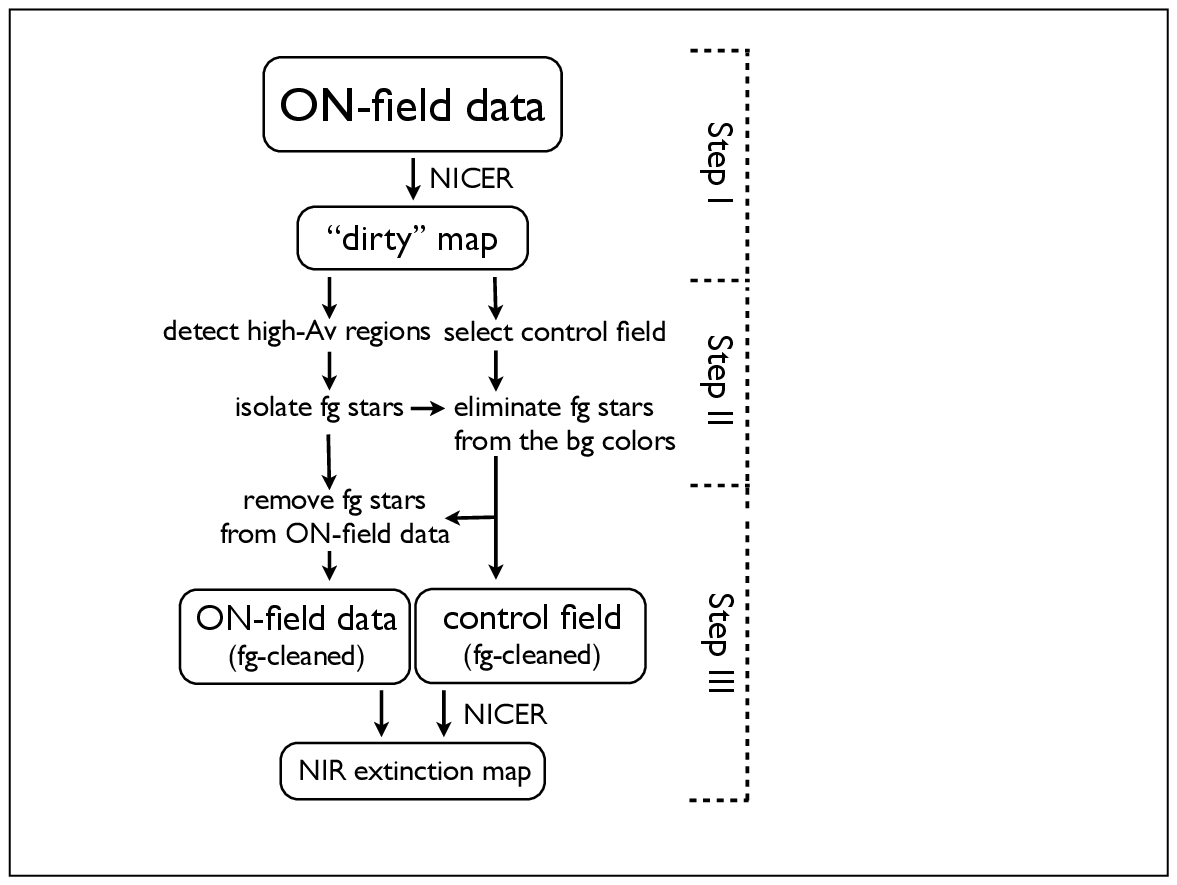} 
      \caption{Flow diagram illustrating the implementation of the \textsf{nicer} technique in this paper (explained in Section \S\ref{subsec:nicer}). The acronyms \emph{fg} and \emph{bg} refer to foreground and background, respectively.
              }
         \label{fig:flow}
   \end{figure}

We retrieved from the UKIDSS archive the $JHK_S$ data for the chosen complexes. For the NIR reddening-law, we adopted the coefficients from \citet{car89}:
\begin{equation}	
\tau_K = 0.600 \times \tau_H = 0.404 \times \tau_J = 0.114 \times \tau_V.
\label{eq:reddening-law}
\end{equation}
The extinction mapping procedure itself consisted of three steps which were required especially to deal with the large number of foreground sources. This procedure is illustrated as a flow-chart in Fig. \ref{fig:flow}. In the first step, a "dirty" extinction map was calculated in order to find a low-extinction region to be used as a control field for the next step of the mapping. The map was also used to determine high extinction regions from which the surface number density of foreground stars, needed in the next step of the map derivation, could be determined.

\begin{figure*}
   \centering
   \includegraphics[bb = 15 15 390 380, clip=true, width=0.325\textwidth]{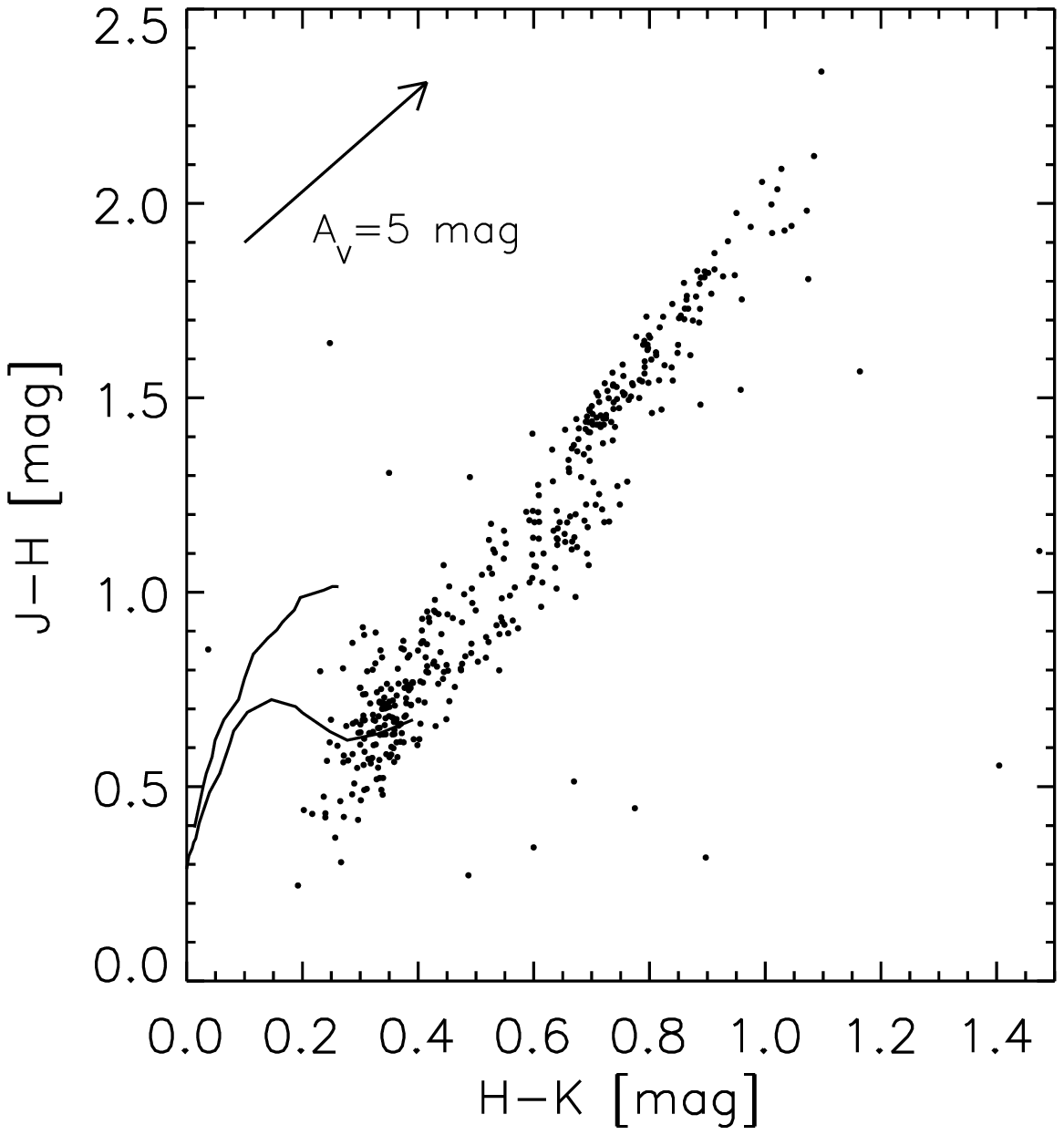} \includegraphics[bb = 70 75 625 625, clip=true, width=0.315\textwidth]{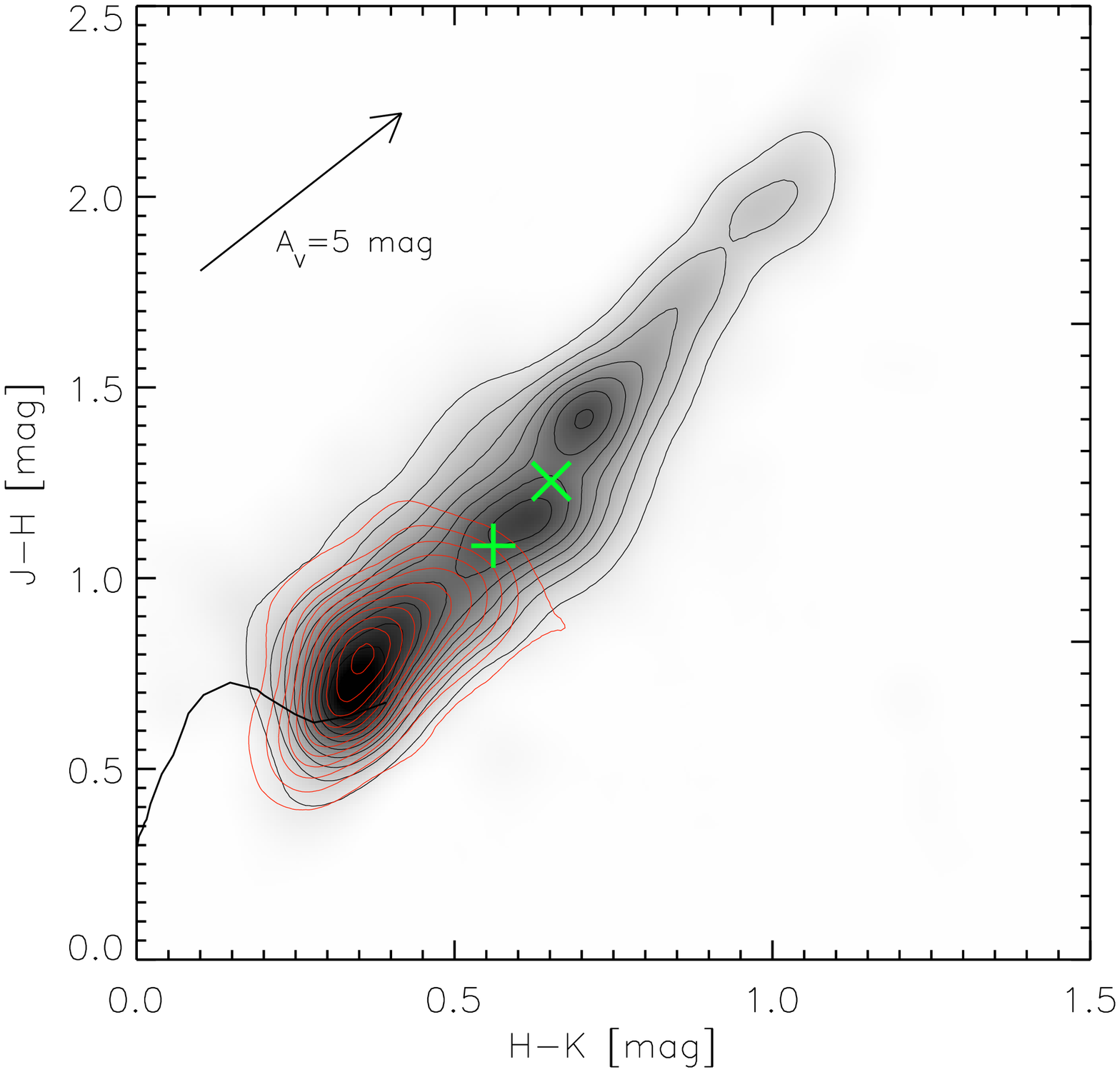}    \includegraphics[bb = 30 10 380 370, clip=true, width=0.3\textwidth]{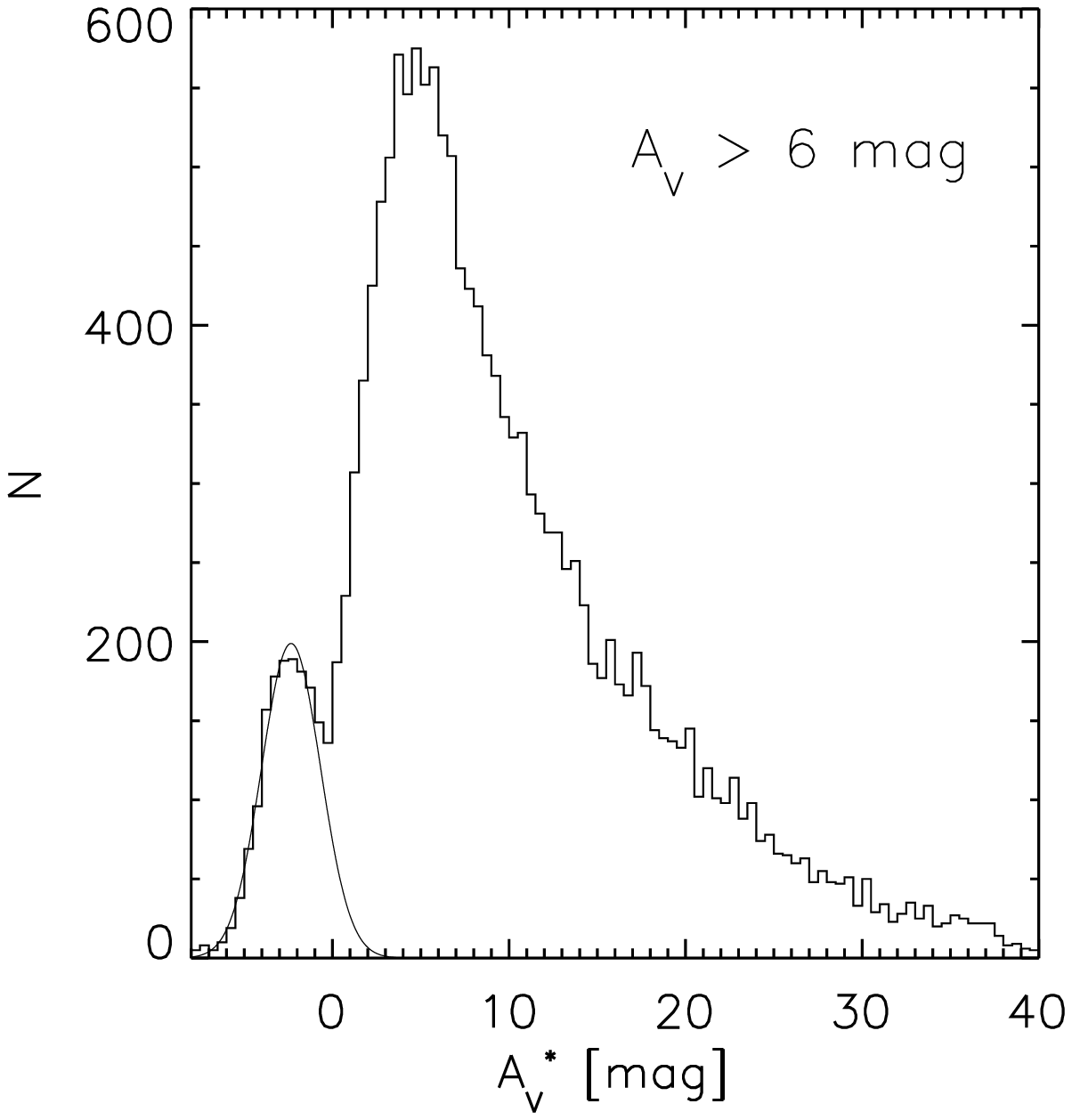}
      \caption{\textbf{Left: }Color-color diagram of the sources in the control field chosen for the 38.94-00.46 complex with photometric error $\sigma < 0.03$ mag. Extinction of $A_\mathrm{V}=5$ mag is indicated with an arrow. \textbf{Center: }The same for all sources in the control field, shown with a grey-scale (and black contours) that indicates the density of points. The mean colors of these sources is marked with a green plus sign. The red contours show the density of the sources that are on the foreground to the 38.94-00.46 complex. The contribution of these sources to the mean color is statistically subtracted from the mean, resulting to the mean color of the background sources that is shown with a green cross. \textbf{Right:} Frequency distribution of the extinction measurements towards individual sources ($A_\mathrm{V}^*$) in a region where $A_\mathrm{V} > 6$ mag. The small bump at slightly negative extinction values are the foreground stars that are removed during the last step of the extinction mapping procedure (see Section \S\ref{subsec:nicer}). The surface density of the foreground sources identified in this way is also used in estimating the distances to the cloud complexes in Section \S\ref{sec:distances}.
              }
         \label{fig:reffield}
   \end{figure*}

In the second step, the contribution of foreground stars was statistically subtracted from the mean colors of the control field in order to calculate the mean color of stars \emph{on the background} to the IRDC complex. In the statistical subtraction, the density of the foreground stars \emph{in JHK color-color space} was first subtracted from the density of all control field stars to yield a foreground corrected source density in each color-color bin. Then, the source number in each color-color bin of the control field was reduced to correspond to this corrected value. The foreground corrected mean colors were then computed using this reduced set of control field stars. The subtraction is important, because the control field is not a truly non-extincted field, but suffers from extinction due to the diffuse, extended dust component along the line of sight in the Galactic plane. This diffuse extinction at the plane varies between $A_\mathrm{V} \approx1-4$ mag kpc$^{-1}$ \citep[e.g.,][]{mar06}. As a result, the stars \emph{farther} away than an IRDC complex are expected to be clearly redder than those \emph{closer} than it. In order to estimate the reddening caused by the IRDC complex itself, it is then necessary to estimate the color of the background population by eliminating the contribution of the foreground sources from the mean colors of the control field. The reddening of the sources in the control field is illustrated in Fig. \ref{fig:reffield}, left panel, which shows the NIR color-color diagram of the sources in the control field chosen for 38.94-00.46. For better visibility, only sources with the photometric errors $\sigma \lesssim 0.3$ mag are plotted. The figure clearly shows how a large fraction of sources is heavily reddened along the reddening line.

The effect of removing the contribution of the foreground stars from the control field is demonstrated in Fig. \ref{fig:reffield} center panel. The figure shows the color-color diagram of all sources in the control field of the 38.94-00.46 complex with a greyscale. The greyscale represents the number density of stars in each color-color bin. The mean colors calculated for this distribution is shown with a green plus sign. The red contours in the figure show the number density of the foreground stars identified from the 38.94-00.46 cloud region (identification procedure is explained in more detail later). The colors of foreground stars are clearly less reddened, as expected for stars that on average are closer to the observer and thus suffer less from diffuse extinction. Using the surface number density of these foreground stars, we subtracted their statistical contribution to the mean colors. This resulted in estimates of the mean colors of \emph{the population on the background to the cloud complex}, indicated in the figure with a green cross. This change in mean color was significant for all mapped clouds, and clearly required for a reasonable definition of the zero-point in the extinction mapping technique. 

The variability of the diffuse extinction component (and the uncertainty in determining the foreground population in the control field) induces uncertainty to the zero-point of the extinction determination. We assessed the level of this uncertainty by performing for one cloud the zero-point determination using 10 different control fields. The standard deviation of the derived zero-points was about $A_\mathrm{V} = 1$ mag. We note that such assessment was not possible for all clouds (or for numerous control fields for one cloud), since finding non-extincted control fields close to the complexes is a problem in general. We consider the result of this experiment to be indicative of the zero-point uncertainty in our method.

In the third and final step the extinction map was corrected for the contribution of foreground sources in the field. The identification (and removal) of foreground sources is trivial in regions where the extinction is high compared to the scatter in the intrinsic colors of stars. The uncertainty of the extinction determination for an individual source is approximately $\sigma (A_\mathrm{V}^*) \approx 3$ mag, so the sources that are located in a region of intermediate local mean of extinction ($ 9 \mathrm{\ mag}\lesssim A_\mathrm{V} \lesssim 20 \mathrm{\ mag}$ mag) can be trivially removed. The effect of foreground stars can be eliminated also at lower extinctions by estimating the surface number density of foreground objects from high $A_\mathrm{V}$ regions and subtracting the estimated contribution of the foreground population to the local mean colors. This procedure is explained in the following.

We estimated the surface number density of the foreground population by examining the sources in regions where the local mean extinction is higher than $A_\mathrm{V} \gtrsim 6$ mag (about 2-$\sigma$ error of an individual extinction measurement). Fig. \ref{fig:reffield} right panel illustrates this procedure by showing the frequency distribution of individual extinction measurements, $A_\mathrm{V}^*$, towards such regions of the 38.94-00.46 complex where the local mean extinction is above $A_\mathrm{V} > 6$ mag. Clearly, the foreground stars form a separate 'bump' in the distribution slightly below $A_\mathrm{V} = 0$ mag. The surface number density of the foreground sources was estimated by fitting a Gaussian to this bump and integrating the number of stars within it. 

In principle, it would be possible to use the determined foreground source density to statistically subtract the foreground stars from the observed field down to $A_\mathrm{V} = 0$ mag. In practice, however, doing this would impose an assumption that all extinction features are caused by a dust component at one well-defined distance. In order to avoid this assumption, we only performed the elimination of the foreground stars down to $A_\mathrm{V}$ regime in which the foreground star density could be determined, i.e. down to such $A_\mathrm{V}$ values in which the foreground bump could be identified from a frequency distribution such as shown in Fig. \ref{fig:reffield} right panel. In practice, this requirement resulted in elimination of foreground stars above $A_\mathrm{V} \gtrsim 4$ mag. Thus, the extinction maps above this limit are, to first order, free from foreground star contamination, but suffer from it below it. However, we note that below this limit the foreground star contamination actually becomes clearly lower, because the relative amount of the background sources is high. 


The spatial resolution of the extinction data is set by the width of the Gaussian smoothing function that is used in smoothing the pencil-beam extinction measurements towards stars ($A_\mathrm{V}^*$) onto a map grid ($A_\mathrm{V}$). In practice, the choice for this is guided by the surface density of the background sources, as all map pixels towards which extinction is to be calculated should have at least one actual measurement (source) inside. For the depth of the UKIDSS data, we concluded that the spatial resolution of $30\arcsec$ represents a good balance between higher resolution and lower noise at intermediate column densities. This translates to the physical resolution of 0.4 pc at 2.5 kpc distance. Typically, the pixels with $A_\mathrm{V} = 20$ mag have a few sources inside the beam area (depends on the Galactic coordinates).

\subsection{CO data}
\label{subsec:codata}

We used the publicly available $^{13}$CO emission line data from the GRS as another tracer of gas in the mapped IRDC complexes. These data provide a tool to independently verify the physical connection between clouds located close to each other in projection. The GRS survey provides $^{13}$CO data for the galactic longitudes $l = [18{\degr}, 55{\fdg}7]$, thus including 8 of the 10 complexes of our sample. The spectral resolution of the data is 0.2 km s$^{-1}$, the typical rms noise 0.4 K, and the spatial resolution is $46\arcsec$. From these data, \citet{rat09} identified structures in position-position-velocity space, resulting to a catalogue of 829 molecular clouds containing 6124 clumps. We use the designations of the GRS catalog as listed in \citet{rom10} to refer to the clouds of our sample (see Table \ref{tab:clouds}). 

We constructed position-velocity diagrams for each region that was mapped in NIR extinction and used them to disentangle the gas component primarily responsible for the extinction features (see Fig. \ref{fig:showcase} for an example). This component was then chosen from the spectral line cubes and integrated to yield the total emission from the cloud complex. The maps of integrated $^{13}$CO intensity are shown for each complex in Appendix \ref{app:figures} (see also Fig. \ref{fig:showcase}). In almost all chosen complexes, the dust extinction features clearly match the morphology of the CO data at the chosen velocity interval, thereby indicating that the extinction features indeed are caused by the gas component in the chosen velocity interval. The correlation between NIR extinction and CO is further discussed in \S\ref{subsec:irdcs_nir}.

\subsection{Mid-infrared opacity data}
\label{subsec:mirdata}

We compared the NIR extinction data also to the mid-infrared dust opacity maps derived from 8 $\mu$m shadowing features in the images of the Spitzer/GLIMPSE survey by \citet{per09}. While that technique reaches higher extinctions (up to $A_\mathrm{V} \lesssim 100$ mag) in high resolution ($\sim 1\arcsec$) compared to NIR mapping, it is considerably less sensitive to low extinction values ($A_\mathrm{V} \approx 1-10$ mag). Therefore, the two techniques complement each other well, with the NIR data probing lower density envelopes and the 8 $\mu$m data probing higher density cores. 

For each IRDC complex, we retrieved the 8 $\mu$m opacity maps\footnote{Accessible via http://www.darkclouds.org/} of all individual IRDCs identified in the region by \citet{per09}. Typically, there were several tens of these IRDCs within the mapped regions. From the maps of individual IRDCs, we then constructed 8 $\mu$m opacity maps for the entire complexes. These maps are shown alongside the NIR extinction maps and the $^{13}$CO integrated emission maps in Fig. \ref{fig:showcase} and in Appendix \ref{app:figures}. In general, the correspondence between the NIR and MIR extinction data is rather scarce, due to both very different spatial resolution and sensitivity. The data are compared more closely in \S\ref{subsec:irdcs_nir}. 

    \begin{figure*}
   \centering
   \includegraphics[bb = 40 5 900 415, clip=true, width=\textwidth]{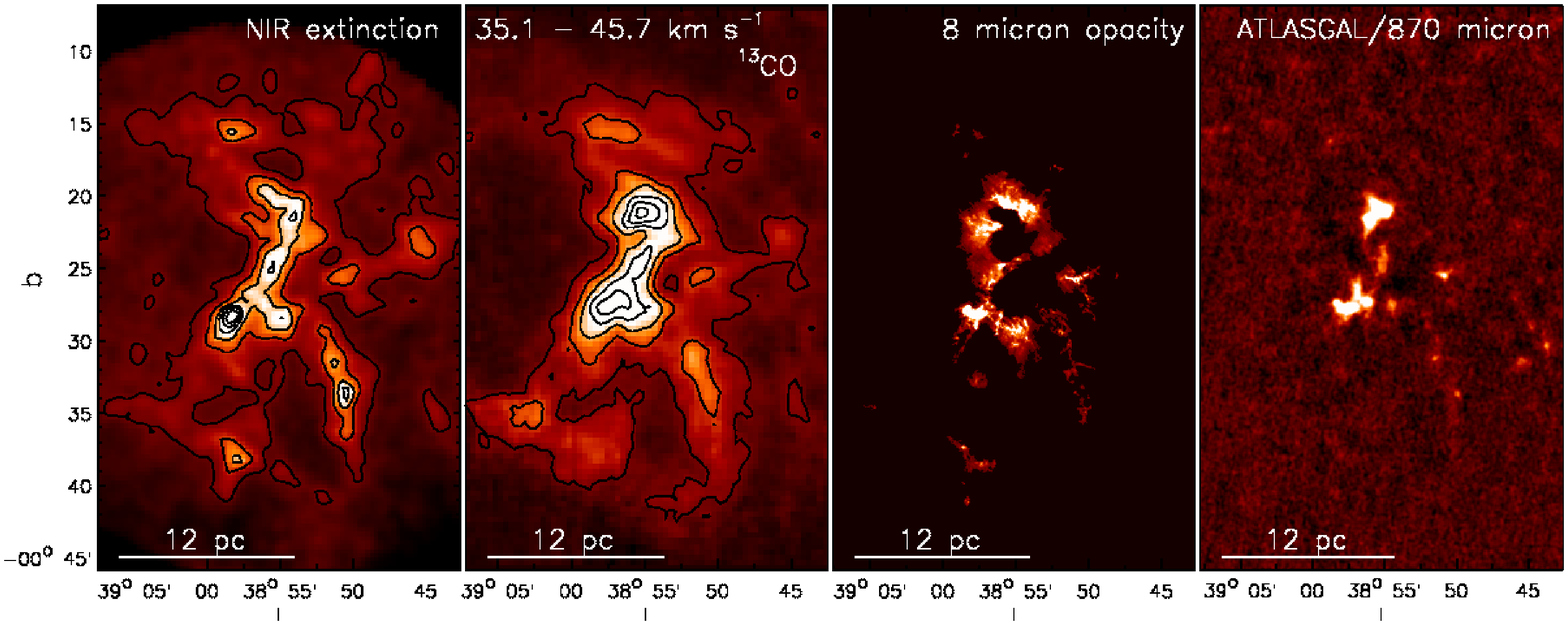}
   \includegraphics[width=0.35\textwidth]{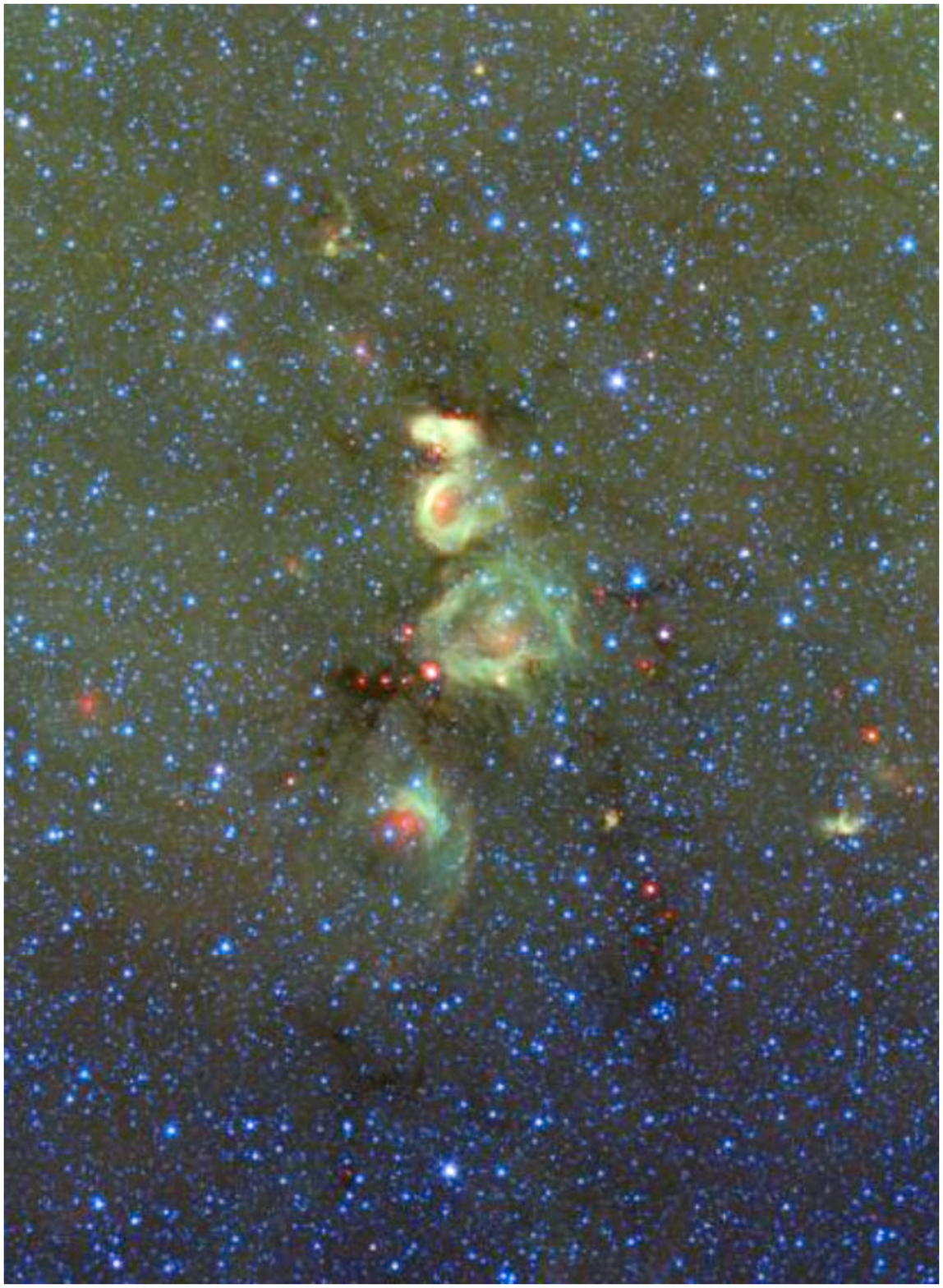}  \includegraphics[bb = 200 0 500 450, clip=true, width=0.315\textwidth]{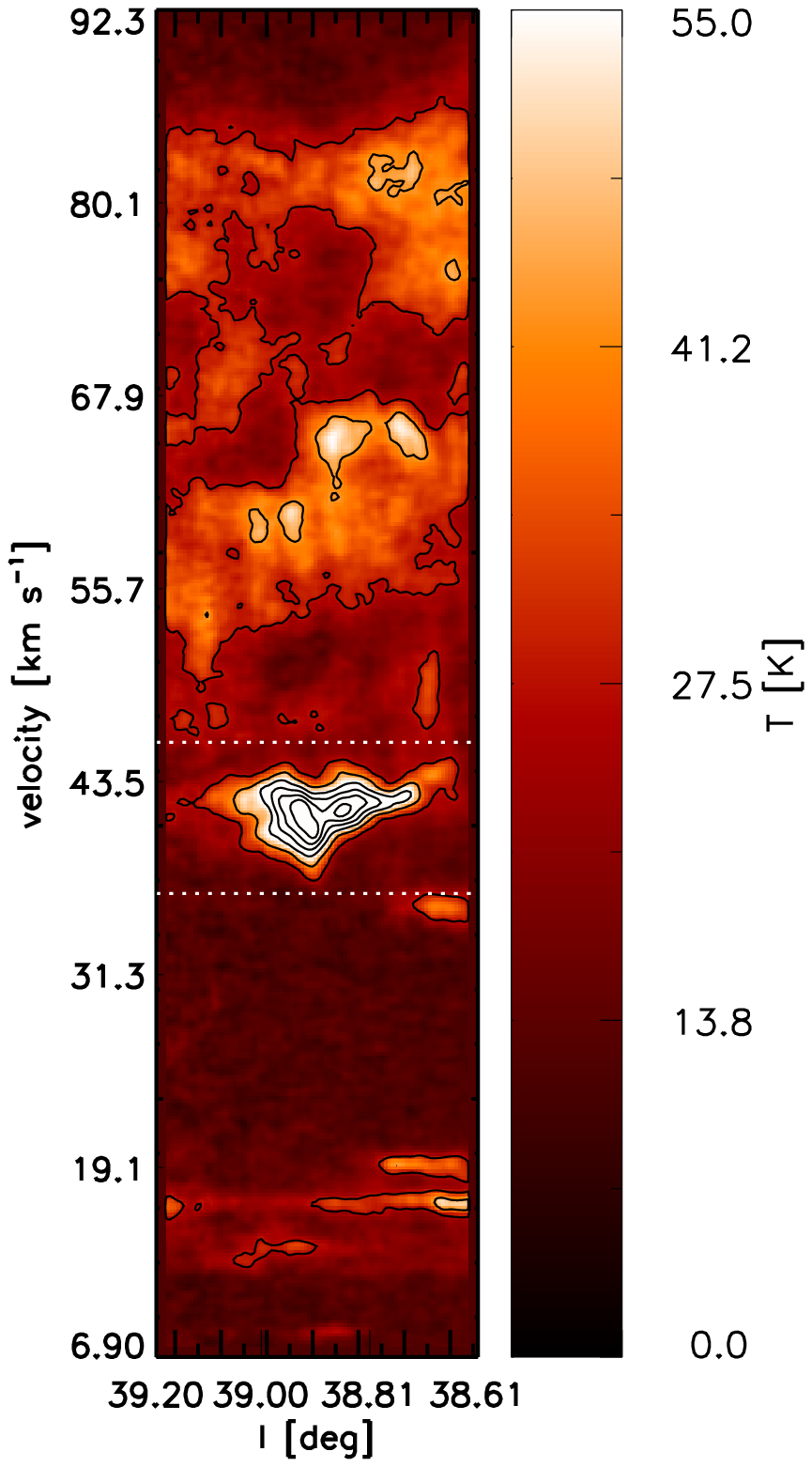}     
   \caption{Infrared Dark Cloud complex 38.94-00.46 at the distance of $D_\mathrm{fg} = 2.7$ kpc (estimated from foreground source density, see Section \S\ref{sec:distances}). \textbf{Top row: }The first panel shows NIR dust extinction map, derived using $JHK_\mathrm{S}$ data from the UKIDSS Galactic Plane Survey. The resolution of the map is $30\arcsec$ and contours starting from $A_\mathrm{V}=3$ mag in steps of 4 mag are shown. The second panel shows the integrated intensity of the $^{13}$CO molecule from the GRS survey data. The contours start from 15 K km s$^{-1}$ and the step is 15 K km s$^{-1}$. The velocity region included in this map is 35.1-45.7 km s$^{-1}$. The third panel shows 8 $\mu$m optical depth map of the complex, constructed from the opacity maps of $\sim 20$ individual IRDCs identified in this region by \citet{per09}. The color scale is linear, starts from $A_\mathrm{V} \approx 10$ mag, and the maximum is $A_\mathrm{V} \approx 90$ mag. The fourth panel shows the 870 $\mu$m dust emission from the region mapped by the ATLASGAL survey \citep{sch09}. The color scale is linear and starts at $A_\mathrm{V} \approx 3.6 $ mag, and the maximum is $A_\mathrm{V} \approx 43 $ mag. \textbf{Lower Left: }Spitzer/GLIMPSE/MIPSGAL 3-color image of the complex, showing 4.5, 8, and 24 $\mu$m data with blue, green, and red, respectively. \textbf{Lower Right: } Position-velocity diagram of the region, integrated over $-00^\circ 45\arcmin<  b < 0^\circ$. The velocity interval chosen to represent the complex is indicated with dotted white lines.   
              } 
         \label{fig:showcase}
   \end{figure*}

    \begin{figure*}
   \centering
   \includegraphics[width=\textwidth]{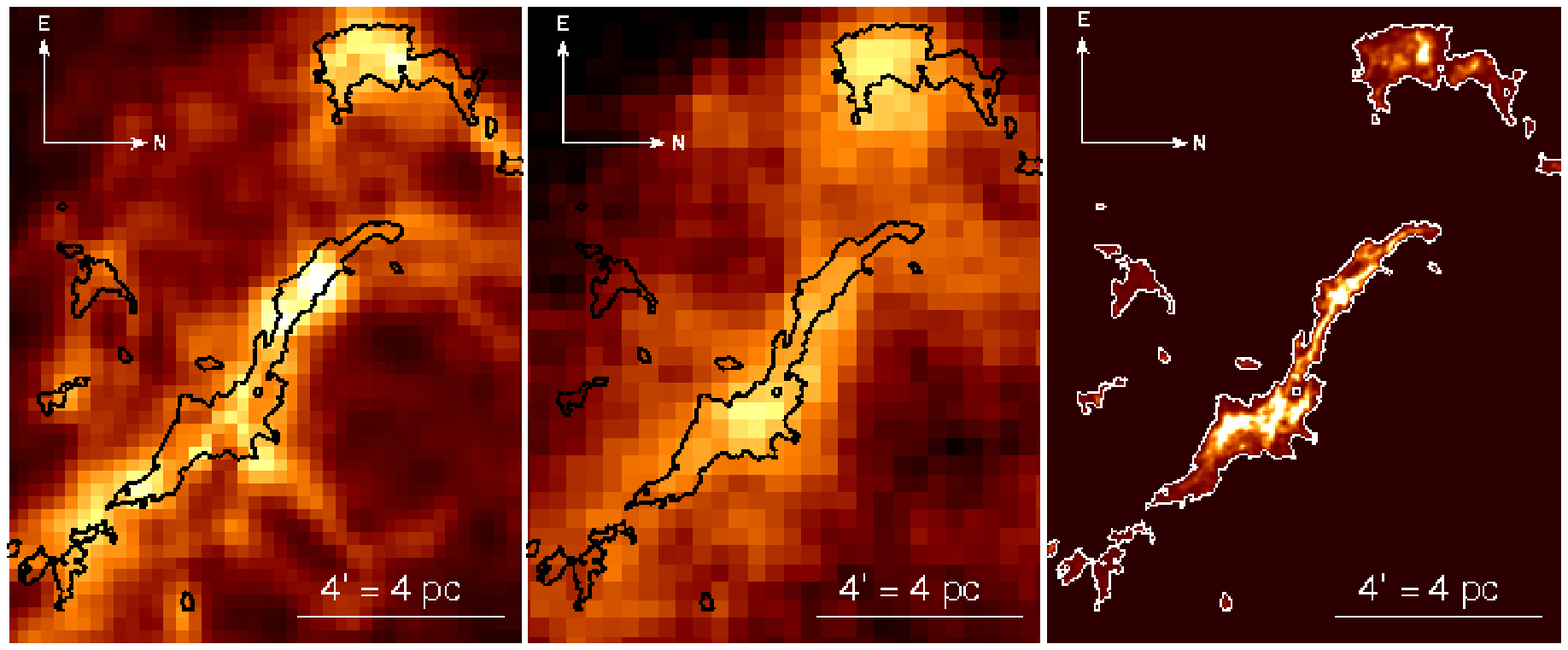}
   \caption{Close-up of a few IRDCs in the 35.49-00.31 complex (see Fig. \ref{fig:35_5} for the map of the entire complex). \textbf{Left: }Near-infrared dust extinction map. The color scale is linear and spans $A_\mathrm{V} = 3-30$ mag. The contour shows the detection threshold of the 8 $\mu\mathrm{m}$ opacity data (see the right panel). \textbf{Center: }Integrated intensity of the $^{13}$CO molecule in the velocity interval $24.6-46.9$ km s$^{-1}$. \textbf{Right: }8 $\mu$m optical depth map. The white contour is drawn at $\tau (8$ $\mu\mathrm{m}) = 0.3$, or $A_\mathrm{V} \approx 6.3$ mag, representing the detection threshold of the data \citep{per09}. The color scale is linear with the maximum of $A_\mathrm{V} = 90$ mag.
              }
         \label{fig:showcase2}
   \end{figure*}

\subsection{870 $\mu$m dust emission data}
\label{subsec:atlasgal}

Thermal dust emission from the Galactic plane ($| l | < 60^\circ $, $|b| < 1.5^\circ$) has been mapped at 870 $\mu$m by the Atacama Pathfinder Experiment (APEX) within the ATLASGAL survey \citep{sch09}. The survey provides emission data in resolution of $19.2\arcsec$ with typical sensitivity of 0.25 Jy/beam (5$\sigma$). This sensitivity limit translates to $N(\mathrm{H}_2) \approx 6.1 \times 10^{21} $ cm$^{-2}$, or $A_\mathrm{V} \approx 6.1$ mag, using typical values of {\bf $T = 20$ K} \citep[e.g.,][]{ega98} and $\kappa_{870} = 1.85$ g cm$^{-2}$ \citep{oss94}. The thermal dust emission observations will not "saturate" at high column densities, unlike NIR and MIR extinction techniques, and thereby they provide perfect complimentary data to trace the densest parts of the IRDC complexes. The data from ATLASGAL for each complex are shown in Figs. \ref{fig:showcase} and Appendix \ref{app:figures} to allow comparison with the other data. In \ref{subsec:masses} we use these data to estimate the mass of dense gas in the complexes.

\section{Distance determination using NIR source counts}
\label{sec:distances}


Nine out of ten complexes in our sample are covered by the GRS survey and kinematic distances have been derived for them by \citet{rat06} and \citet{rom09}. Using a different approach, \citet{mar09} used NIR data from the 2MASS survey together with the Besan\c{c}on Galactic stellar distribution model to derive distances towards a large number of IRDCs, including six complexes of our sample. We list the distances from both sources in Table \ref{tab:clouds} using notation $D_\mathrm{grs}$ for kinematic distance measurements and $D_\mathrm{gen}$ for \citet{mar09} distances. Similarly with the technique adopted by \citet{mar09}, the photometric data from the UKIDSS/GPS employed in this paper can be used to determine cloud distances. With the help of the derived extinction maps, it is straightforward to measure the surface density of stars that are located between the cloud and the observer, i.e. \emph{on the foreground} to the cloud. The measured surface density can then be compared to what is predicted for the Galactic coordinates of the cloud by the stellar distribution model of the Galaxy, which yields an estimate for the cloud distance \citep[e.g.,][]{alv98}. This technique have been demonstrated to yield distance estimates that are in agreement with estimates made with other techniques for nearby molecular clouds \citep[][]{lad09, lom11}. We used this approach to estimate the distance for the clouds in the sample. In the following, the distance determination procedure is explained in detail.


For each mapped complex, we estimated the surface density of the foreground stars by counting the sources in $K$ band in regions where the derived extinction was between $5-8 \mathrm{\ mag} < A_\mathrm{V} < 20$ mag, depending slightly on the Galactic coordinates of the cloud (because the error of the extinction measurement depends on the coordinates). In these regions, the foreground stars are easily distinguished from background stars due to their low extinction values compared to the local mean extinction (illustrated in Fig. \ref{fig:reffield} and explained in \ref{subsec:nicer}). For the distance determination, it is preferable to count the foreground stars down to as faint magnitude as possible in order to keep the uncertainty due to counting statistics (Poisson error) as low as possible. Yet, the sources should only be counted down to the completeness limit of the photometric data, $K_\mathrm{lim}$. Since the completeness limit of the UKIDSS/GPS data changes as a function of the Galactic coordinates, the limiting magnitude above which the source counts were performed changed from cloud to cloud. Typically, the sources brighter than $K_\mathrm{lim} < 16.5$ mag were included. Typically, the regions from which the sources were counted were $\sim 50$ arcmin$^2$ ($\sim 800$ map pixels) and contained $\sim 1000$ foreground sources. 

\begin{figure}
   \centering
   \includegraphics[bb = 25 10 380 370, clip=true, width=0.5\columnwidth]{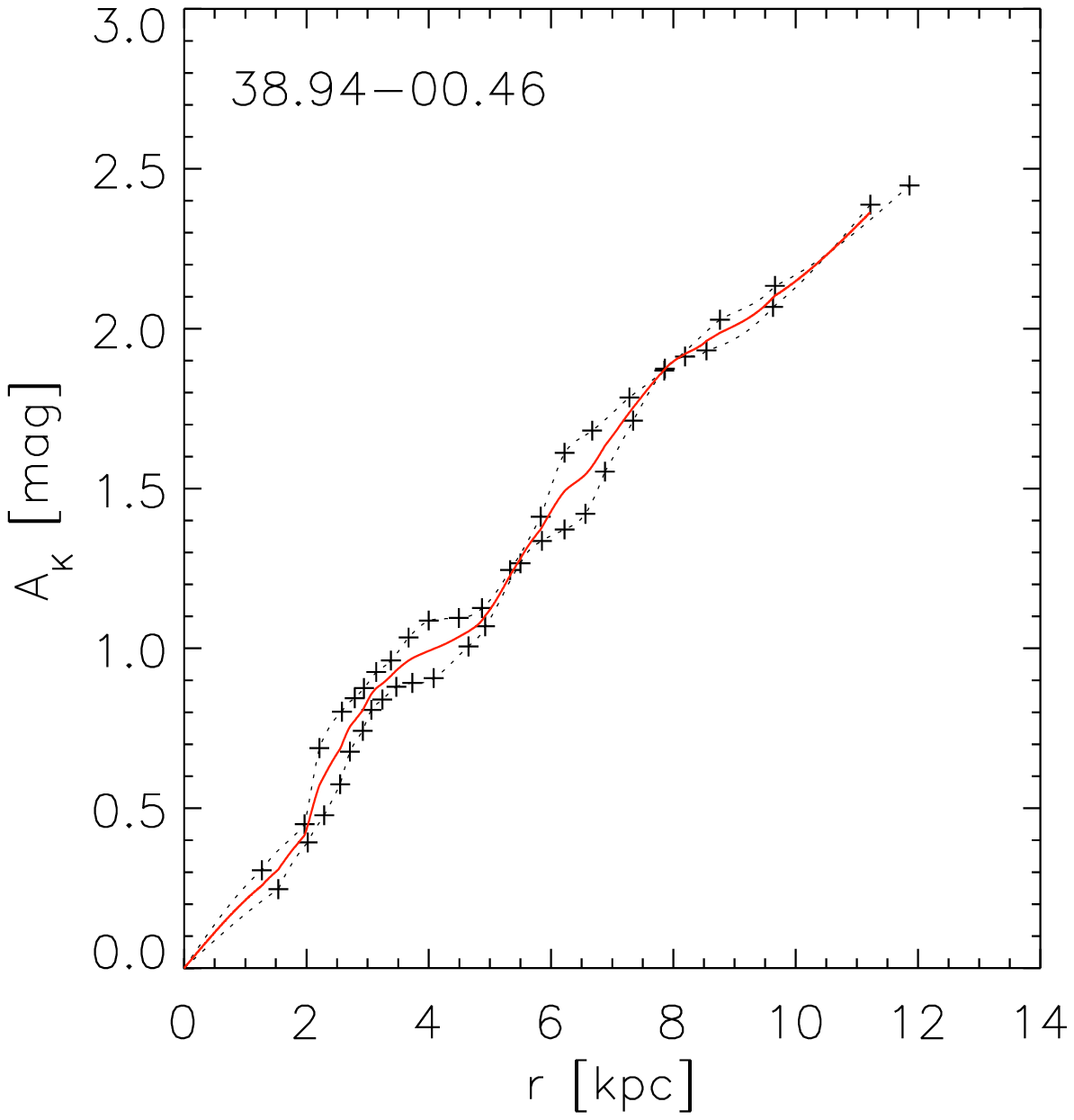}\includegraphics[bb = 25 10 380 370, clip=true, width=0.5\columnwidth]{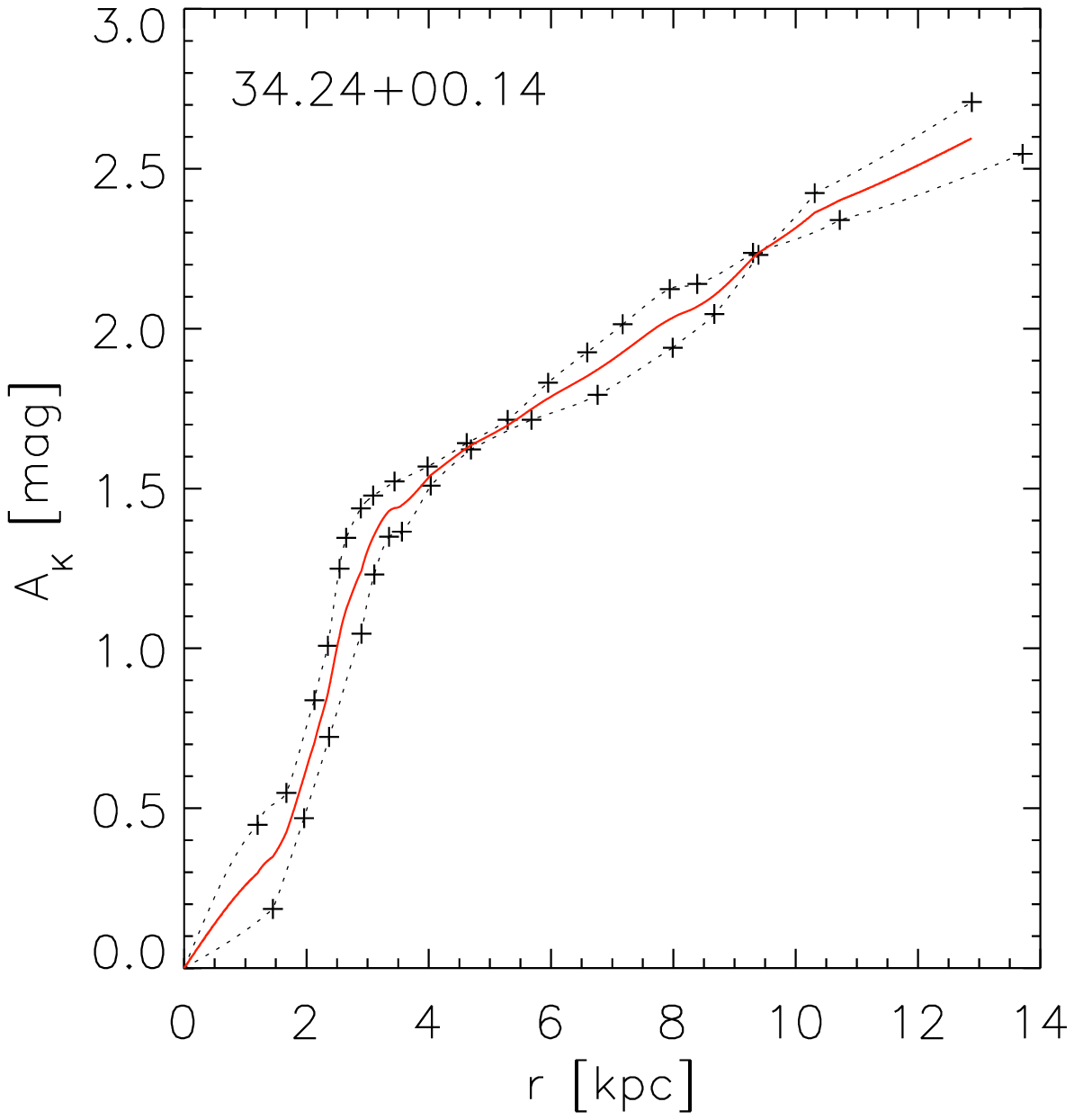}
   \includegraphics[bb = 25 10 500 350, clip=true, width=0.5\columnwidth]{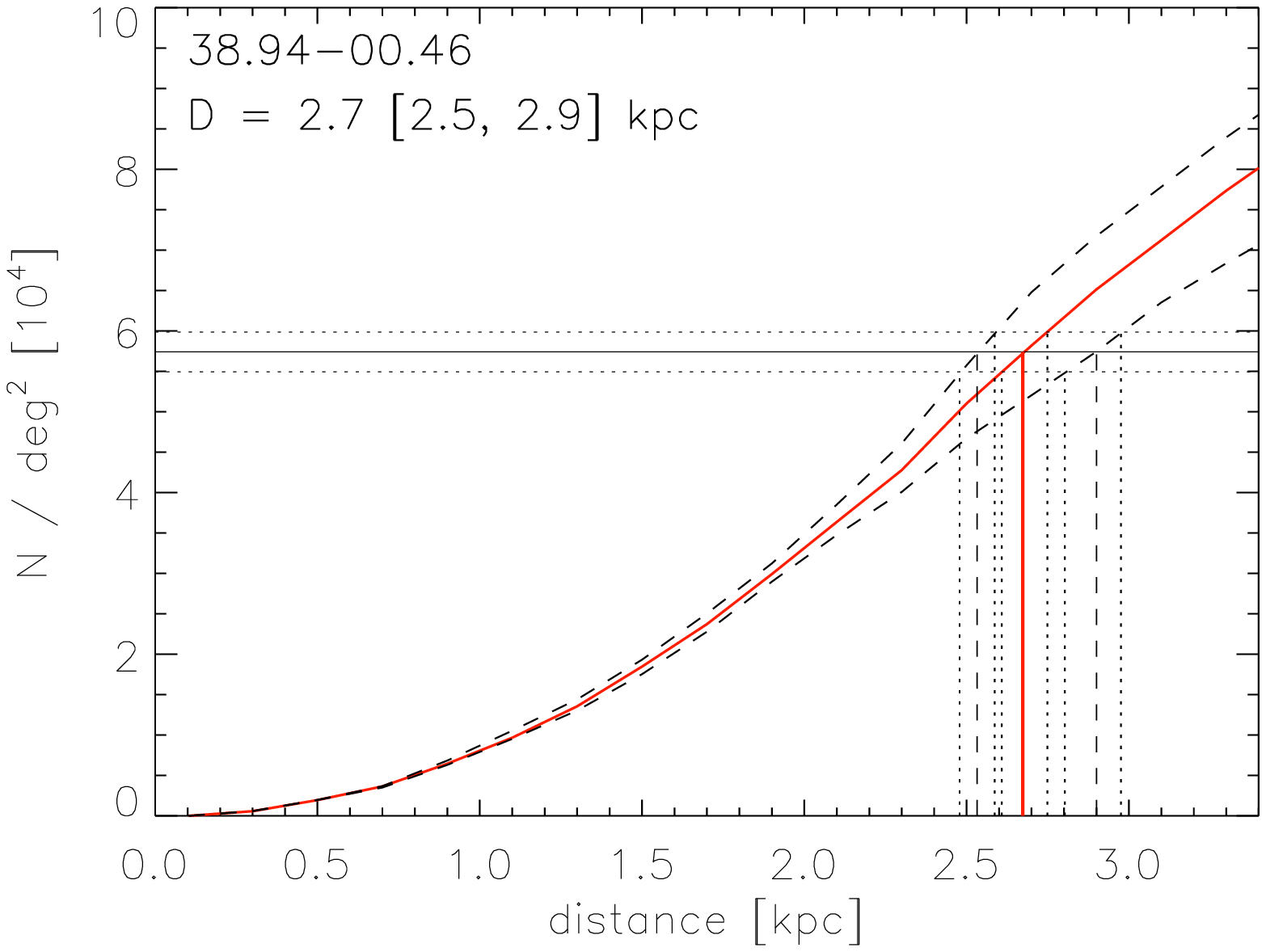}\includegraphics[bb = 25 10 500 350, clip=true, width=0.5\columnwidth]{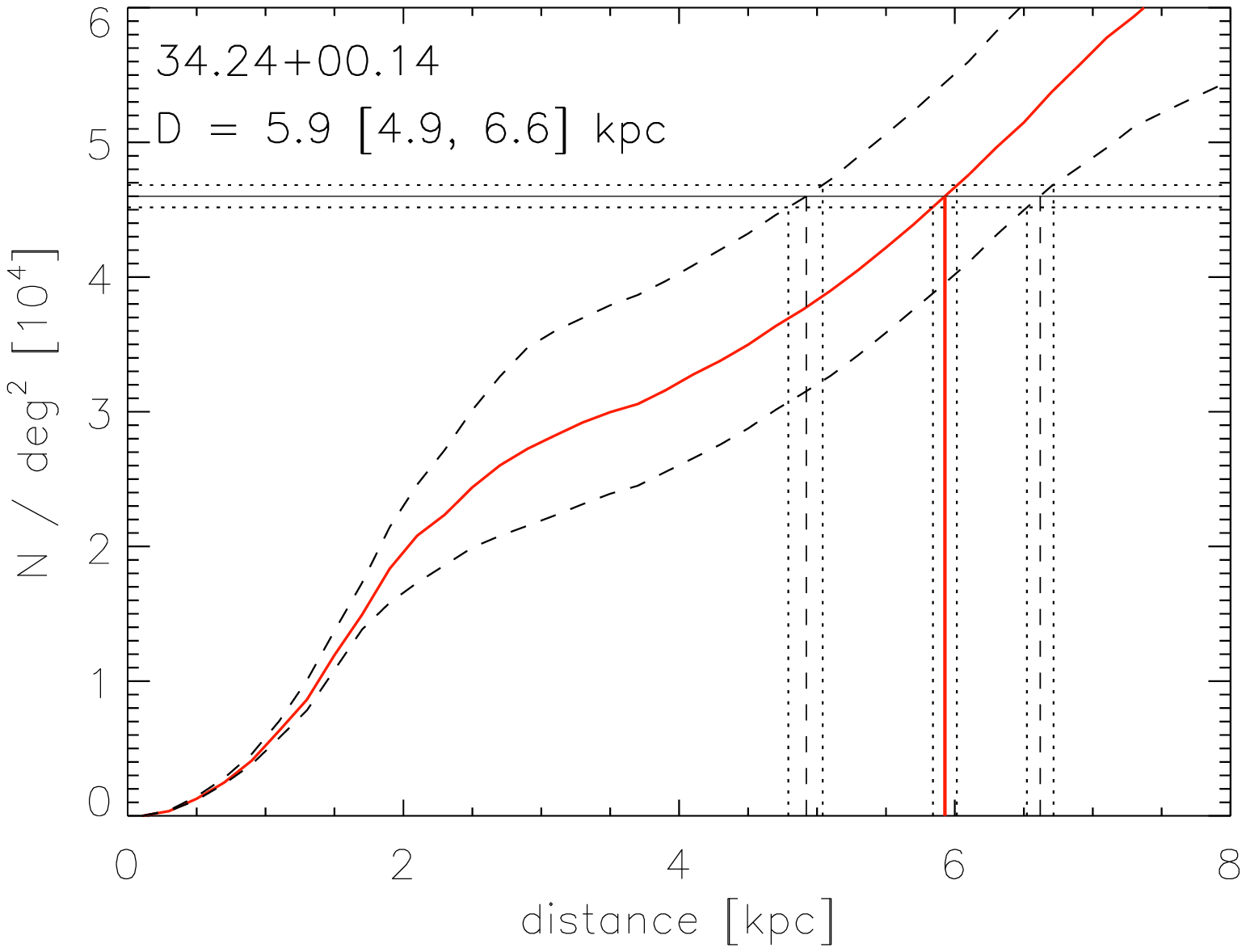}   
   \caption{{\bf Top row: }Radial dust extinction profiles constructed for two complexes. The plus signs and the dotted lines show the extinction profiles adopted from \citet{mar06}. The solid red line shows the mean profile. {\bf Bottom row: }Predicted source count functions (source count above the limiting magnitude $K_\mathrm{lim}$), based on the Besan\c{c}on model and \citet{mar06} reddening data, for the same two clouds. The source count functions corresponding to the mean extinction profile are shown with the solid red curves. The dashed black curves show the same function using the individual extinction profiles. The horizontal lines show the observed foreground source density and the vertical lines point out the corresponding distances. The dotted lines show the Poisson errors related to the observed source density (and the resulting distances).}
         \label{fig:distance}
\end{figure}

The observed foreground source densities were then compared to the source counts predicted by the Besan\c{c}on Galactic stellar distribution model\footnote{Available on-line at http://model.obs-besancon.fr/ .} \citep{rob03}. The model was used to retrieve differential source counts in the distance bins of 0.2 kpc, down to the previously chosen limiting magnitude $K_\mathrm{lim}$, at the positions of the mapped clouds. For the purpose of distance determination, the most crucial parameter in the stellar distribution model is the amount of the diffuse, inter-cloud extinction from which all sources, including the foreground sources, suffer. This diffuse extinction generally amounts to $\sim$1-4 mag kpc$^{-1}$ and can vary rapidly as a function of Galactic position \citep[e.g.,][]{mar06}. In particular, \citet{mar06} used the 2MASS NIR data to derive radial extinction profiles in the Galactic plane in spatial intervals of 15' and radial bins of  $\sim 200-500$ pc. We used these data to estimate the level of diffuse extinction towards the IRDC complexes. This was done by choosing 2-4 positions surrounding the complex (depending on the morphology of the complex) and computing the mean  extinction profiles for those lines of sight. As an example, Fig. \ref{fig:distance} shows profiles for two clouds, one showing a very linear diffuse extinction component (38.94-00.46) and another showing a more complicated diffuse component (34.24+00.14). The figure shows two individual profiles towards these clouds with plus signs and dotted lines, and the mean of these profiles is shown with a red solid line. 


The differential source counts retrieved from the Besan\c{c}on model were then modified (reddened) using the mean diffuse extinction profiles shown in Fig. \ref{fig:distance}. These reddened count data were  used to compute the final predicted source densities as a function of distance. Then finally, the observed foreground source densities were compared to this function to yield the distance estimate. Fig. \ref{fig:distance} illustrates the predicted cumulative source count functions as a function of distance. The source count function corresponding to the mean extinction profile is drawn with a red solid line. The observed foreground source density is indicated with a solid horizontal line, and the associated Poisson counting error that results from the finite region available for the source density determination is shown with dotted lines. This statistical error is typically $\approx 100$ pc. The resulting distance is shown with solid vertical lines. We refer in this paper to the distances derived from the foreground source densities with $D_\mathrm{fg}$ and list the distances for all clouds in Table \ref{tab:clouds}.

Figure \ref{fig:distance} also shows the predicted source count functions computed using two individual extinction curves available for the cloud (dashed curves). This illustrates the uncertainty in the distance determination due to the uncertainty in the radial extinction profile. For complexes with almost linear extinction profiles, the scatter in the resulting distances is typically 5-10 \%. For complexes showing more complex profiles it can amount up to $\sim 15$ \%. This clearly dominates the error of the distance measurement over the statistical uncertainty. We conclude that the uncertainty of our distance determination is about 15 \%.


\section{Results}           
\label{sec:results}

\subsection{IRDC complexes as viewed by NIR extinction}
\label{subsec:irdcs_nir}


We used the technique described in \S\ref{subsec:nicer} to derive dust extinction maps for the 10 cloud complexes listed in Table \ref{tab:clouds}. Figure \ref{fig:showcase} shows, as an example, the map derived for the complex $38.94-00.46$, together with the $^{13}$CO integrated line emission, the 8 $\mu$m dust opacity, and 870 $\mu$m dust emission maps. We give in Appendix \ref{app:figures} the same figures for all complexes of the sample. In addition, Fig. \ref{fig:showcase2} shows a smaller-scale zoom-in to the central region of the complex 35.49-00.31 to demonstrate the relatively high resolution of the NIR extinction data. 


In general, the NIR extinction maps reveal extended (tens-of-parsec size), highly structured cloud complexes in the mapped regions. In the chosen resolution ($30\arcsec$, corresponding to 0.44 pc at 3 kpc distance), the maps probe extinction values up to $A_\mathrm{V}\lesssim 30-40$ mag. The maps commonly reveal how the IRDCs (identified from 8 $\mu$m data) in the regions are connected by lower column density ($A_\mathrm{V} \sim 5$ mag) structures. While these low-column density structures already strongly suggest a physical connection between the IRDCs in the regions, the connection is confirmed by the $^{13}$CO data showing the structures to be connected in the position-position-velocity space. 

\begin{figure}
   \centering
   \includegraphics[bb = 70 75 625 625, clip=true, width=0.528\columnwidth]{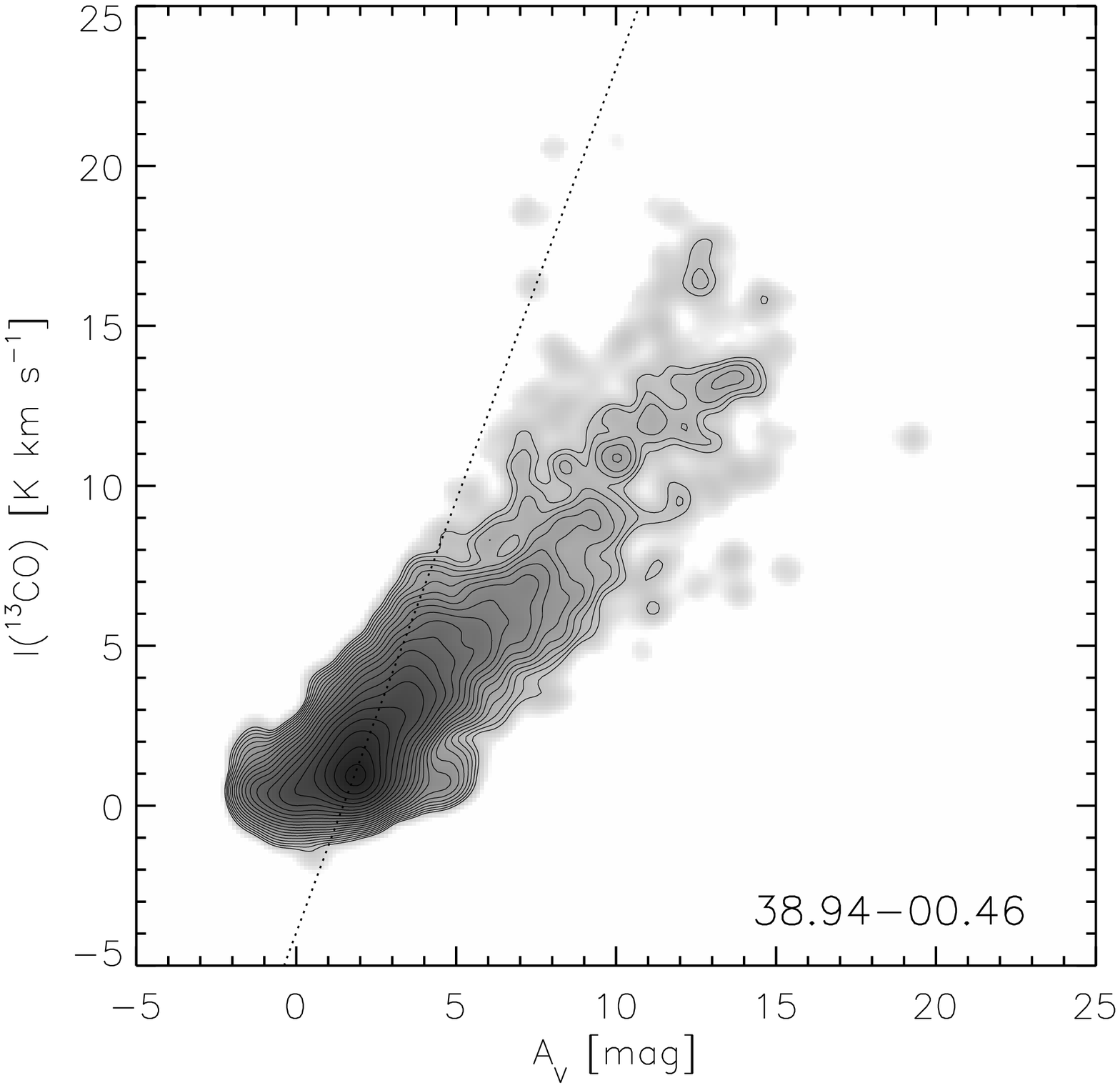}\includegraphics[bb = 130 75 625 625, clip=true, width=0.472\columnwidth]{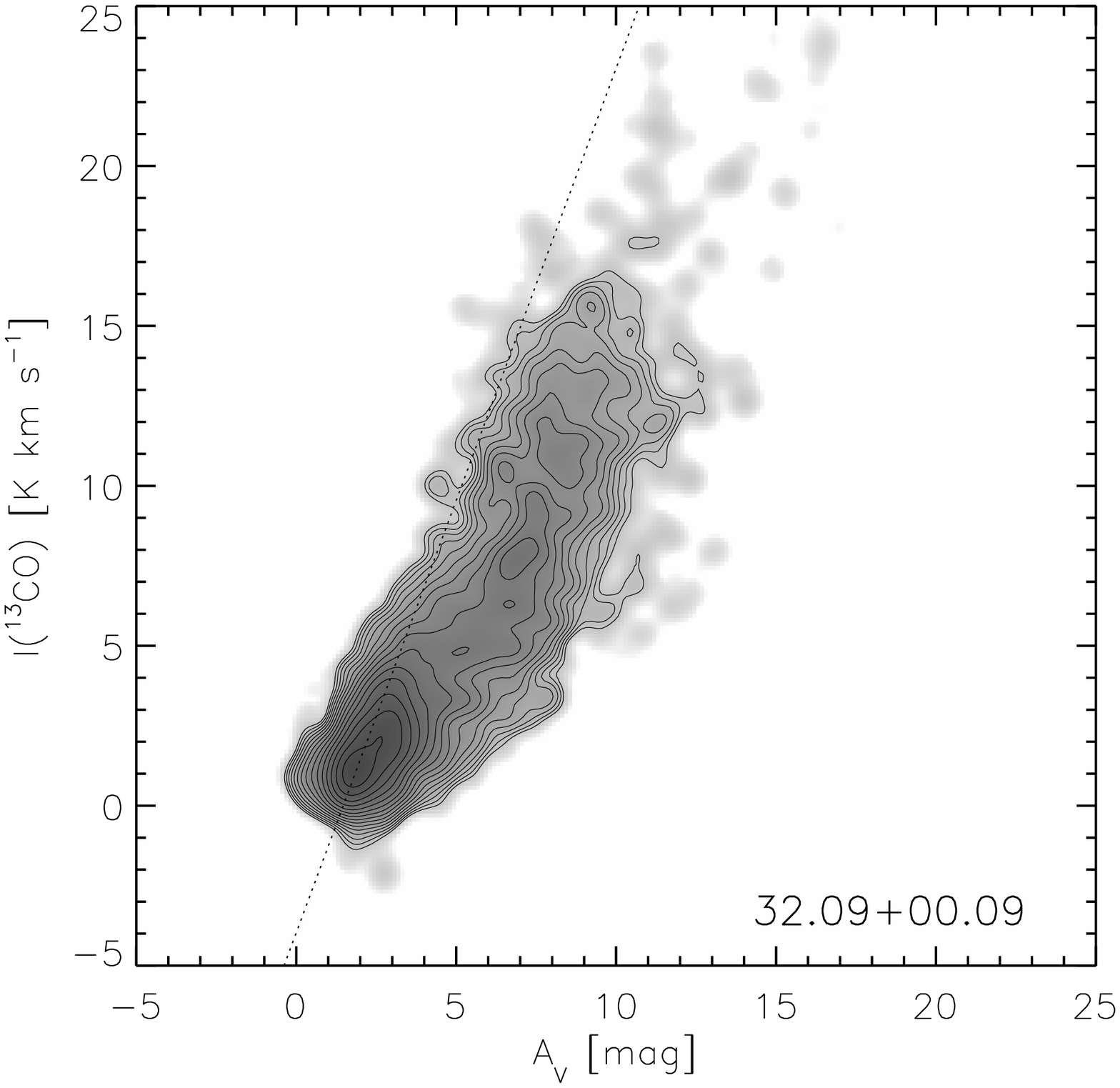} 
      \caption{Pixel-to-pixel comparison of NIR extinction and $^{13}$CO integrated line emission for the 38.94-00.46 ($D_\mathrm{fg} = 2.7$ kpc) and 32.09+00.09 ($D_\mathrm{fg} = 8.0$ kpc) complexes. The dotted line shows a typical ratio of these quantities for local molecular clouds \citep{pin08}. 
              }
         \label{fig:pixel-to-pixel}
   \end{figure}

Generally, the NIR extinction features resemble closely the morphology of the $^{13}$CO emission in the chosen velocity intervals. To illustrate the correspondence, Fig. \ref{fig:pixel-to-pixel} shows a pixel-to-pixel comparison of the NIR and $^{13}$CO data for one of the closest clouds (38.94-00.46, $D=2.7$ kpc) and for the farthest cloud (32.09+00.09, $D=8.0$ kpc). The comparison shows a clear correlation between the quantities. In most cases, as for example in the case of 38.94-00.46 shown in Fig. \ref{fig:pixel-to-pixel}, the CO emission saturates around $A_\mathrm{V}\approx 4-6$ mag, a feature commonly observed in CO-to-A$_\mathrm{V}$ ratios of molecular clouds and associated to optical depth effects and/or depletion of CO at higher column densities \citep[e.g.,][]{com91, pin08}. The figure also shows the ratio for these quantities measured typically at low column densities in local molecular clouds \citep[in particular, for the Perseus molecular cloud,][]{pin08}. At low column densities, the observed data roughly agree with this relation.


The NIR extinction data and 8 $\mu$m opacity data have a factor of $\sim 15$ difference in spatial resolution and $\sim 5$ in sensitivity, which makes comparing them directly with each other difficult. Qualitatively, the morphology seen in the NIR extinction (and CO) maps is usually recognizable also in the 8 $\mu$m cloud detections. This is clearly shown in, e.g., Fig. \ref{fig:showcase} for the complex 38.94-00.46 and in Fig. \ref{fig:11_1} for 11.10-00.10. This correlation is no surprise, because the complexes for the study were chosen as high contrast features from the Spitzer/GLIMPSE images (as described in \S\ref{subsec:sample}). Again qualitatively, the morphologies of smaller-scale structures, i.e., individual IRDCs, vary tremendously between the NIR and 8 $\mu$m data. While in many cases no IRDCs are detected at the positions of NIR extinction peaks (and vice versa), in other cases the correlation is very good. As an example, Fig. \ref{fig:showcase2} shows a zoom-in to the central region of the 35.49-00.31 complex. The 8 $\mu$m, NIR and CO detections in this region correlate very well. 


It is worth noting that unlike 8 $\mu$m data the NIR extinction mapping is relatively unaffected by the signatures of star-forming activity in the clouds. An example of this is shown in Fig. \ref{fig:showcase}, in which a Spitzer/GLIMPSE/MIPSGAL 3-color image of 38.94-00.46 is shown\footnote{Available through http://www.alienearths.org/glimpse/ .}. The cloud region contains several nebulae that are likely excited by the formation of intermediate- or high-mass stars. Clearly, these nebulae affect strongly the MIR opacity map, practically suppressing the detection of the MIR shadowing features (''holes'' in the 8 $\mu$m opacity map). However, the nebulae are not as  strong sources in the NIR and thereby do not strongly affect the NIR extinction map. 

\subsection{Masses of the complexes from different tracers}
\label{subsec:masses}


The NIR extinction data can be used to estimate the total gas column density along the line of sight by first converting the NIR extinction to visual extinction (Eq. \ref{eq:reddening-law}), and then converting the the visual extinction to the total hydrogen column density using the measured ratio \citep{boh78}
\begin{equation}
2N(\mathrm{H}_2)+N(\mathrm{H}) = 1.9 \times 10^{20} \times A_\mathrm{V} \mathrm{\ [cm}^{-2} \mathrm{\ mag}^{-1}\mathrm{]} .
\label{eq:bohlin}
\end{equation}
To calculate the total masses from the column densities, we summed up all the extinction values in the clouds above $A_\mathrm{V} = 2.5$ mag. This relatively high threshold was chosen to better disentangle the cloud complexes from the surrounding, possibly physically unrelated extinction components. We used the distances for the clouds derived from the foreground stellar density measurements (see Sec. \ref{sec:distances}), $D_\mathrm{fg}$, that are listed in Table \ref{tab:clouds} in calculating the masses. The resulting masses are listed in Table \ref{tab:clouds} and they are generally between 1 and $10 \times 10^{4}$ M$_\odot$. For comparison, the total masses of the nearby Taurus and Orion A molecular clouds above $A_\mathrm{V} = 1$ mag are $\sim 10^4$ M$_\odot$ and $\sim 10^5$ M$_\odot$, respectively \citep[e.g.,][]{kai09b}. However, a large fraction of the total mass of the cloud is likely in regions below our selected threshold of $A_\mathrm{V} > 2.5$ mag. This fraction is generally $\sim 40-80$ \%, and for example $\sim 43$ \% in Taurus  and Orion A \citep[e.g.,][]{kai09b, lom10}. Thus, most complexes included in this study have masses on the order of $\sim 10^5$ M$_\odot$, equaling roughly the mass of the Orion A molecular cloud. We note that the largest uncertainty in the derived masses originates from the distance uncertainty. For example, a distance uncertainty of 500 pc for a cloud at 4 kpc distance results in $\sim 30$ \% uncertainty in total mass. The uncertainty of the total masses induced by the uncertainty of the extinction zero-point determination ($A_\mathrm{V} \approx 1$ mag, see Section \S\ref{subsec:nicer}) is on the order of 15 \%, resulting from the fact that typical mean extinctions above the mass calculation threshold are $<A_\mathrm{V}> \approx 6-7$ mag.


For comparison, we also list in Table \ref{tab:clouds} the total masses of the clouds calculated from the GRS CO line emission data by \citep{rom10}, corrected to correspond to the distance that was used in calculating the masses from NIR data. In the \citet{rom10} work, the cloud masses were calculated using $^{13}$CO line emission. The excitation temperature was estimated from the $^{12}$CO line assuming that it is optically thick. This temperature was adopted for calculation of $^{13}$CO optical depths and thereby $^{13}$CO column densities. Finally, the H$_2$ column densities were estimated from the $^{13}$CO column densities using the relations $n(^{12} \mathrm{CO} ) / n(^{13}\mathrm{CO}) = 45$ and $n(^{12}\mathrm{CO}) / n(\mathrm{H}_2) = 8 \times 10^{-5}$ \citep{lan90, bla87}. The masses calculated from CO emission are relatively close to the masses derived from NIR extinction, although they seem to be systematically larger by a factor of $\sim 1.5-2$. This difference likely has its origin in the different techniques used to identify clouds from data sets. \citet{rom10} employes thresholding in position-position-velocity space using the CLUMPFIND algorithm by \citet{wil94}, while we identified clouds with a simple column density threshold in the plane of the sky. The difference in the total masses may also be induced by the fact that the zero-point determination of the extinction mapping is not very robust for clouds at large distances (see Section \S\ref{subsec:nicer} for discussion).


We also calculated the total masses of the mapped regions from the 8 $\mu$m opacity data. For the transformation between opacity and column density, we adopted the reddening curve between 8 $\mu$m and NIR as given by \citet{ind05}, i.e. $\tau_{8\mu\mathrm{m}} = 0.43 \times \tau_\mathrm{K}$. With the reddening curve between NIR and visual from \citet{car89}, this yields the conversion factor $\tau_{8\mu\mathrm{m}} = 0.048 \times A_\mathrm{V}$. Then, Eq. \ref{eq:bohlin} was again used to transform $A_\mathrm{V}$ to hydrogen column densities. The resulting masses are listed in Table \ref{tab:clouds}. The masses from 8 $\mu$m emission are typically on the order of $\sim 10^4$ M$_\odot$, i.e. about an order of magnitude smaller than the NIR masses. This is quite expected, because the 8 $\mu$m data probe (only) significantly higher column densities than NIR data, thereby not detecting the bulk of the cloud mass. Obviously, the 8 $\mu$m absorption is spatially less extended than NIR or CO. This comparison of the total masses is given solely to demonstrate the extent of the diffuse component in the clouds not detectable via 8 $\mu$m data.


The masses of the dense gas component in the complexes were also calculated from the 870 $\mu$m dust emission data. In this calculation, we assumed for simplicity a constant temperature of $T = 20$ K for all complexes. Then, assuming optically thin emission, the column densities can be calculated from the equation \citep{sch09}
\begin{equation}
N_{\mathrm{H}_2} = \frac{F_\nu R}{B_\nu (T_d) \Omega \kappa_\nu \mu m_H},
\end{equation}
where $R$ is the gas-to-dust-ratio, $R = 100$, $B_\nu$ is the Planck function at temperature $T_d$, $\Omega$ is the beam size, $\kappa_\nu = 1.85$ g cm$^{-2}$ is the mass absorption coefficient at $870$ $\mu$m, and $\mu = 2.8$ is the mean molecular weight. The resulting masses are given in Table \ref{tab:clouds}. The masses traced by the dust emission are generally on the order {\bf of $\sim 10-20$ \%} of the masses traced by NIR and CO data, and thereby close to the 8 $\mu$m masses. 


\begin{table*}
\begin{minipage}[t]{2.\columnwidth}
\caption{Cloud complexes and their properties.}             
\label{tab:clouds}      
\centering          
\renewcommand{\footnoterule}{}  
\begin{tabular}{l c c c c c c c}
\hline\hline       	
Complex & $D_\mathrm{fg}$\footnote{Distance derived from the foreground source densities in this paper (see Section \ref{sec:distances}).} & $D_\mathrm{gen}$\footnote{Distance derived from modeling of NIR source color distribution by \citet{mar09}.} & $D_\mathrm{grs}$\footnote{Unless otherwise stated, the distance is a kinematic distance from the GRS survey \citep{rom09}.} &  $M_\mathrm{NIR} [10^4$ M$_\odot]$ & $M_\mathrm{CO} / M_\mathrm{NIR}$\footnote{As estimated in \citet{rom10}, corrected to correspond to the distance $D_\mathrm{fg}$ listed in column 2.} & $M_{8 \mu\mathrm{m}} / M_\mathrm{NIR} [\%] $  & $M_{\mathrm{870\ }\mu\mathrm{m}} / M_\mathrm{NIR}$ [\%] \\ 
\hline
$03.80-01.00$	& 2.5 	 	 & -		& -	 	& 				 		1.7		  &  - 	&-				  & 12  \\
$11.10-00.10$	& 4.1 	 	 & 4.74	&  3.6 	& 						4.5		  &  -	   &11			  & 15  \\
$18.54-00.16$	& 5.9	 	 	 & 4.33	&  4.1   	&  						4.7		  &  1.1 &17			  & 22  \\
$24.94-00.16$	& 3.5 		 & -		&  3.4\footnote{\citet{rat06}}	&  		3.6		  &  2.7 &10			  & 10  \\
$32.09+00.09$	& 8.0 	 	 & 7.45 	&  7.1	&  						12 		  &  1.5 &29			  & 26  \\
$34.24+00.14$	& 5.9 	 	 & -		&  3.8	&  						13		  &  2.7 &46			  & 51  \\
$34.39-00.71$	& 0.7 	 	 & -		&  1		&  						0.9		  &  0.3 &0				  & 3 \\
$35.49-00.31$ 	& 3.2 	 	 & 3.63	&  3.0	&  	 					5.6  	  	  &  0.7 &8				  & 7  \\
$36.74-00.16$	& 5.1 		 & 3.88	&  3.6	&				 		5.4		  &  1.7 &7				  & 14 \\
$38.94-00.46$	& 2.7 	 	 & 4.14	&  10.5 	&  						2.0		  &  1.6 &50			  & 14  \\
\hline                  
\end{tabular}
\end{minipage}
\end{table*}

\section{Discussion}  
\label{sec:discussion}


The NIR extinction mapping technique employed in this paper provides a view to molecular clouds harboring potential progenitors of massive stars over a column density range $A_\mathrm{V} \approx 2-40$ mag, with a spatial resolution of $\sim 30\arcsec$. This range greatly complements other commonly used column density tracers, i.e., CO line emission, mid-infrared dust extinction, and thermal dust emission. While CO obviously can provide kinematic information not reachable via NIR data, the dynamical range of NIR data extends to clearly higher column densities and is free from depletion/optical depth effects. This allows the NIR data to probe, in a uniformly calibrated manner, the regime from diffuse envelopes to the denser clumps in the clouds. Compared to other dust tracers, the NIR data is more sensitive to low column densities. It can also provide a reasonable overlap with the dynamical ranges of dust emission and MIR extinction, allowing cross-calibration. Perhaps most importantly, the NIR extinction mapping can provide temperature independent data on the mass distributions of IRDC complexes in spatial resolution that matches that of the Herschel satellite that will provide high sensitivity dust emission data of the Galactic plane \citep[e.g.,][]{mol10}. 


We independently derived the distances for the complexes by comparing the observed foreground stellar densities to the predictions from the Besan\c{c}on stellar distribution model. We estimated the uncertainty of our approach to be about 15 \%, resulting mostly from the variable diffuse extinction component in the Galactic plane. There is a reasonable agreement between the kinematic distance estimates and earlier distance estimates based on modeling stellar distributions using 2MASS data ($D_\mathrm{grs}$ and $D_\mathrm{gen}$ in Table \ref{tab:clouds}, respectively), considering the uncertainties of each method. With only 10 clouds in our sample, we did not quantify the differences in the distances derived with different methods in more detail. Compared to the kinematic distance estimates, the insensitivity of our approach to the \emph{near}/\emph{far} distance ambiguity is an asset. For example, in the case of the 38.94-00.46 complex it seems clear that the kinematic distance determination algorithm used by \citet{rom09} has incorrectly chosen the \emph{far} distance for the cloud. As a downside, the foreground stellar densities can only be used to determine distances to clouds within $\sim 8-10$ kpc distance. In summary, the distance measurements confirmed the NIR extinction mapping technique to be feasible for clouds as far as $\sim 8$ kpc away, thus allowing a vast sample of IRDCs to be mapped with the method. 


Compared to mapping local ($D \lesssim 500$ pc) molecular clouds, applying the NIR dust extinction mapping technique to clouds at the distances of a few kiloparsecs has two main obstacles. First, the large fraction of foreground stars towards the clouds biases the determination of extinction at $A_\mathrm{V} \lesssim 4$ mag. As discussed in Section \S\ref{subsec:nicer}, below this column density the contribution from such stars biases the measurement towards lower $A_\mathrm{V}$ values. 
Second, the distribution of diffuse dust at the Galactic plane can have rapid spatial variations making determining the reddening caused solely by the cloud in question uncertain. These two short-comings have the main effect that the zero-point of the $A_\mathrm{V}$ mapping is clearly less well-defined as in mapping nearby clouds. This can have a significant effect on the estimates such as the total masses of the clouds, since most of the molecular cloud mass is always at low column densities. 


The gas component traced by the NIR extinction matches closely the morphology of the CO line emission, with the difference that it extends to clearly higher column densities (i.e., $A_\mathrm{V} \approx 30-40$ mag). Consequently, the masses of the complexes derived from NIR data are comparable to the ones derived from CO emission. Comparison of these masses with the masses derived from 8 $\mu$m dust opacity and 870 $\mu$m dust emission shows that on average these higher column density tracers reveal $\sim 10-20$ \% of the masses of the clouds inside which they are embedded (with a large scatter). However, as speculated in Section \ref{subsec:masses}, it is likely that our total mass estimate from NIR extinction still significantly underestimates the total masses of the complexes due to the relatively high threshold for mass determination, and thereby the masses revealed by 8 $\mu$m dust opacity and 870 $\mu$m dust emission are even clearly smaller, likely $\lesssim 10$ \% of the total masses.

\begin{figure}
   \centering
\includegraphics[width=\columnwidth]{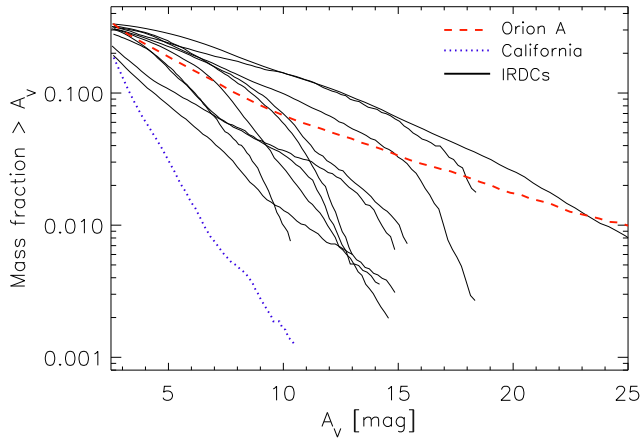} 
      \caption{Cumulative mass functions for the complexes, indicating the fractional mass above a chosen column density (or visual extinction, $A_\mathrm{V}$) threshold. The dotted red and dashed blue lines show the relation for Orion A (star-forming) and California (more quiescent) clouds, respectively, both of which are about $10^5$ M$_\odot$ clouds \citep{kai09b}.
              }
         \label{fig:cmfs}
   \end{figure}

Figure \ref{fig:cmfs} shows the cumulative mass functions calculated from the NIR extinction data. The functions indicate what fraction of the cloud mass is at column densities higher than the reference value (abscissa). For this plot, all the maps were smoothed to common resolution of 0.86 pc, reflecting $30\arcsec$ resolution at the 5.9 kpc distance (except 32.09+00.09 which is at 8.0 kpc distance). For comparison, the figure shows the mass functions of the nearby Orion A and California molecular clouds derived from similar data by \citet{kai09b}. Clearly, the IRDC complexes of our sample contain relatively large amount of high density material. Their mass functions seem to be markedly closer to that of the active star-forming Orion A cloud than the more quiescent California cloud. In particular, none of the IRDCs shows as steep mass function as California. We note that the mass functions presented here are based on data that have $\sim 8$ times lower spatial resolution than data of \citet{kai09b}, and thereby the relations may not be directly comparable. However, we find them clearly indicative, implying that the IRDC complexes show relatively high fraction of high-column density material comparable to very active nearby molecular clouds. 

While recent studies based on the molecular line data of the GRS have generated a census of molecular clouds in the Galactic plane \citep[e.g.,][]{jac06, rat09, rom10}, the dust extinction maps can provide a mass tracer for these clouds extending over a clearly wider dynamical range than the CO data. Most importantly, the dynamical range will overlap with that of higher column density tracers (especially thermal dust emission), allowing combining the information from the two tracers. Such an approach can provide column density data extending from low column density envelope material to gravitationally dominated objects in the clouds. Thereby, the data sets together can be used to reveal the detailed mass distributions surrounding the progenitors of high-mass stars, and we will continue towards deriving such data in forthcoming work. 

\section{Conclusions} 
\label{sec:conclusions}

In this paper, we examined the feasibility of NIR dust extinction mapping technique in tracing low-to-intermediate column density structures surrounding prospective birthplaces of high-mass stars, i.e., IRDCs. We used the data from the UKIDSS/Galactic Plane Survey to derive dust extinction through 10 cloud complexes at the distances between $\sim 2.5-8$ kpc, harboring altogether hundreds of IRDCs. We compared the derived NIR extinction maps to the $^{13}$CO molecular line data from the Boston University-FCRAO Galactic Ring Survey, 8 $\mu$m dust opacity data from a recently published Spitzer IRDC catalogue by \citet{per09}, and to 870 $\mu$m dust emission data from the ATLASGAL survey \citep{sch09}. The conclusions of our work are as follows. 

   \begin{enumerate}

      \item We demonstrated that the NIR dust extinction mapping technique provides a powerful tool to trace the mass distribution of typical IRDC complexes. Using only moderately deep UKIDSS/GPS data ($K \lesssim 19$ mag), extinction maps covering the dynamical range of $A_\mathrm{V} \approx 2 - 40$ mag can be constructed in the resolution of $FWHM \approx 30\arcsec$. Such data provide highly complimentary, \emph{temperature independent} mass distribution data to be combined, e.g., with data soon available from the Herschel satellite mission. Similarly, the results encourage for dedicated, deep NIR observations of IRDC complexes, because the spatial resolution of the extinction data at the high source density of the Galactic plane can be greatly improved with deeper observations.
   
      \item The NIR extinction maps reveal the low column density ($A_\mathrm{V} < 10$ mag) structures surrounding IRDCs in great detail. The IRDCs are dense clumps in these parental (giant) molecular clouds that contain typically $\sim 5-10$ \% of the total masses of the clouds. 

      \item Despite their low star-forming activity the IRDCs of our sample contain relatively large fraction of high column density material, as judged by their cumulative mass distributions. Their mass distributions resemble more closely that of active star-forming clouds (like Orion) than less active clouds (like California). 

     \item We derived distances for the complexes in our sample by comparing the observed NIR foreground source counts to the Besan\c{c}on Galactic stellar distribution model. We estimate the accuracy of the distance measurements to be about 15 \%. The distance measurements confirm the feasibility of the extinction mapping method up to distances of $\sim 8$ kpc.
	
   \end{enumerate}

\begin{acknowledgements}
This publication makes use of molecular line data from the Boston University-FCRAO Galactic Ring Survey (GRS). The GRS is a joint project of Boston University and Five College Radio Astronomy Observatory, funded by the National Science Foundation under grants AST-9800334, AST-0098562, \& AST-0100793. JK acknowledges the Finnish Academy of Science and Letters/V\"ais\"al\"a Foundation in supporting this work.
\end{acknowledgements}



\pagebreak

\appendix

\section{Dust extinction maps and CO data.}
\label{app:figures}

   \begin{figure*}
   \centering
\includegraphics[bb = 155 20 500 700, clip=true, width=0.6\textwidth]{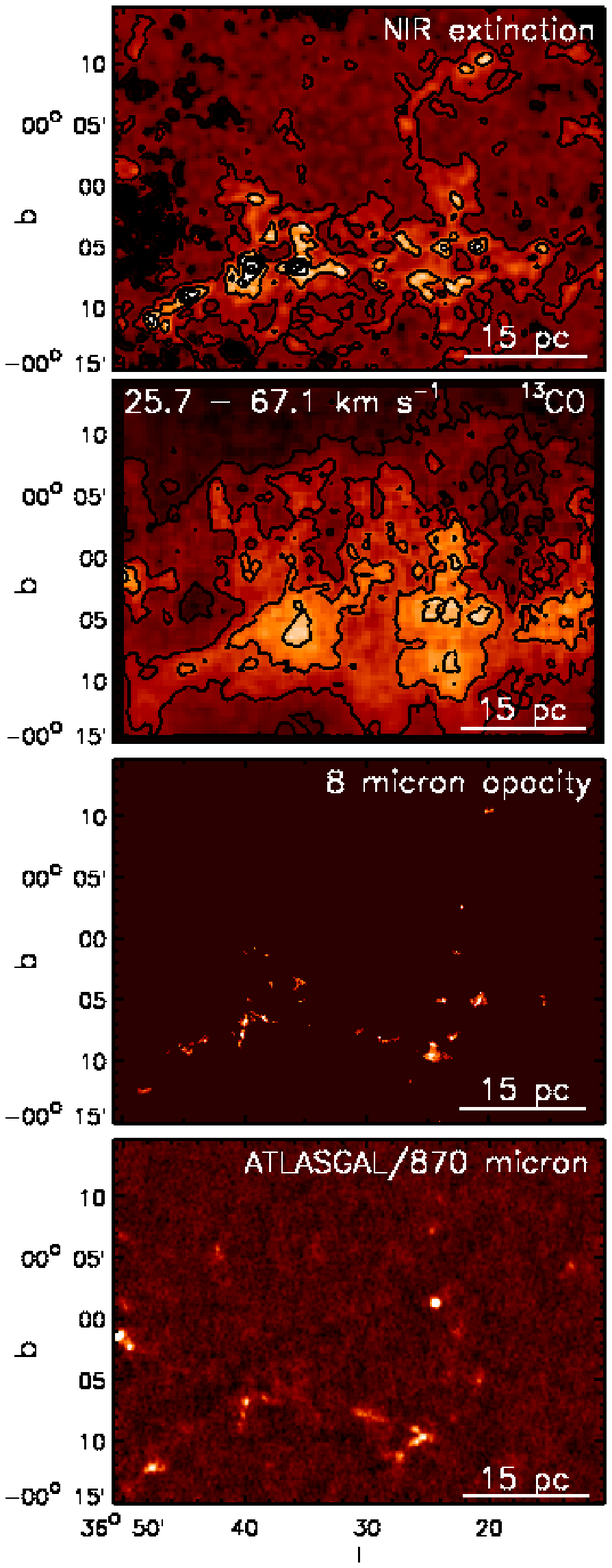}
      \caption{IRDC complex 36.74-00.16. The panels show the NIR dust extinction, $^{13}$CO line emission, 8 $\mu$m dust opacity, and 870 $\mu$m dust emission maps of the complex. The contours of the $A_\mathrm{V}$ map start from 3 mag and the step is 4 mag. The contours of the CO map are in multiples of 15 K km s$^{-1}$. The velocity range chosen for $^{13}$CO data is shown in the CO map. The 8 $\mu$m color scale saturates white at $A_\mathrm{V} \approx 20$ mag and the color scale of the 870 $\mu$m map at $A_\mathrm{V} \approx 7$ mag. The panels show a scale bar corresponding to the distance derived using foreground source count method ($D_\mathrm{fg}$ in Table \ref{tab:clouds}).
              }
         \label{fig:36.6}
   \end{figure*}

   \begin{figure*}
   \centering
\includegraphics[bb = 155 20 500 700, clip=true, width=0.6\textwidth]{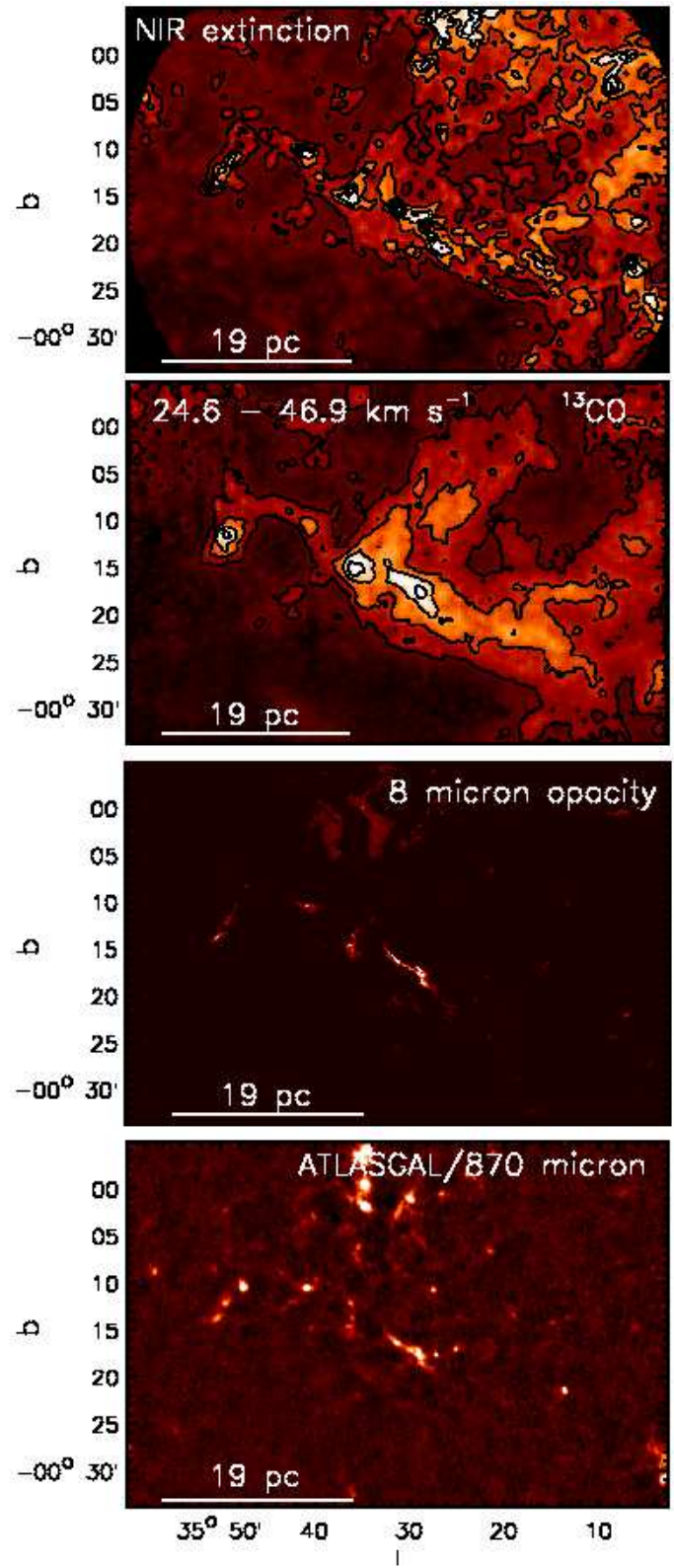}
      \caption{Same as Fig. \ref{fig:36.6}, but for IRDC complex 35.49-00.31. The 8 $\mu$m color scale saturates white at $A_\mathrm{V} \approx 60$ mag and the color scale of the 870 $\mu$m map at $A_\mathrm{V} \approx 15$ mag. 
      See also Fig. \ref{fig:showcase2} for a close-up of the central region of the complex. 
              }
         \label{fig:35_5}
   \end{figure*}

   \begin{figure*}
   \centering
\includegraphics[bb = 155 20 500 700, clip=true, width=0.6\textwidth]{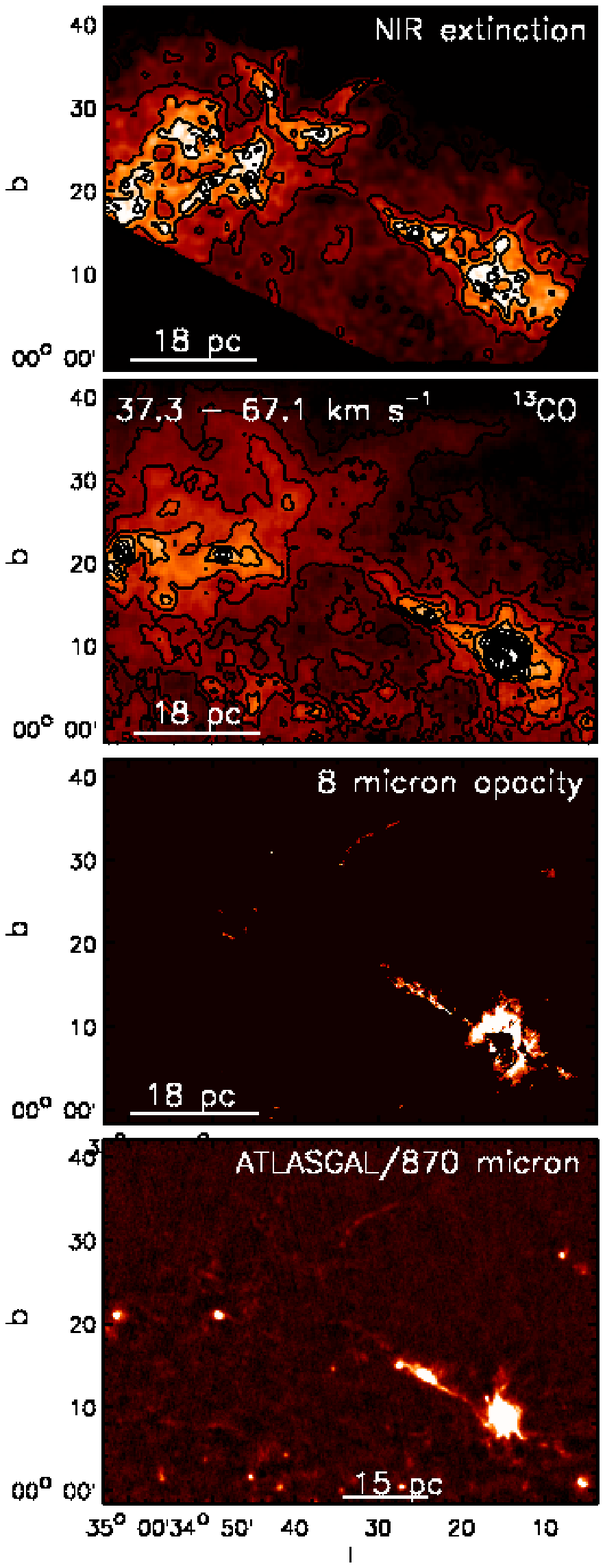}
      \caption{Same as Fig. \ref{fig:36.6}, but for IRDC complex 34.39-00.71. The 8 $\mu$m color scale saturates white at $A_\mathrm{V} \approx 40$ mag and the color scale of the 870 $\mu$m map at $A_\mathrm{V} \approx 7$ mag. 
               }
         \label{fig:34_3}
   \end{figure*}

   \begin{figure*}
   \centering
\includegraphics[width=0.9\textwidth]{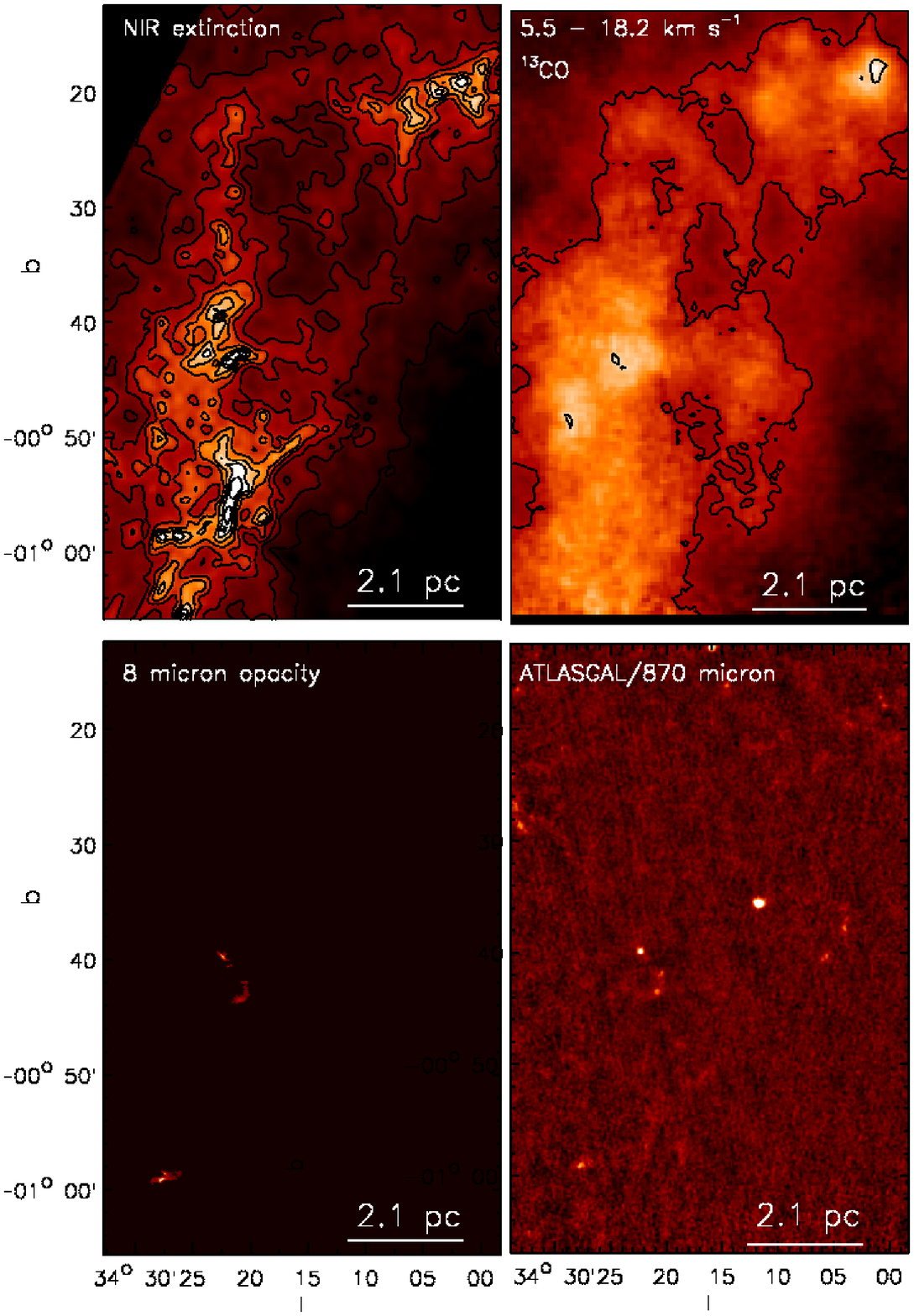}
      \caption{Same as Fig. \ref{fig:36.6}, but for IRDC complex 34.24+00.14. The 8 $\mu$m color scale saturates white at $A_\mathrm{V} \approx 20$ mag and the color scale of the 870 $\mu$m map at $A_\mathrm{V} \approx 11$ mag. 
              }
         \label{fig:34_4}
   \end{figure*}

   \begin{figure*}
   \centering
\includegraphics[bb = 155 20 500 700, clip=true, width=0.6\textwidth]{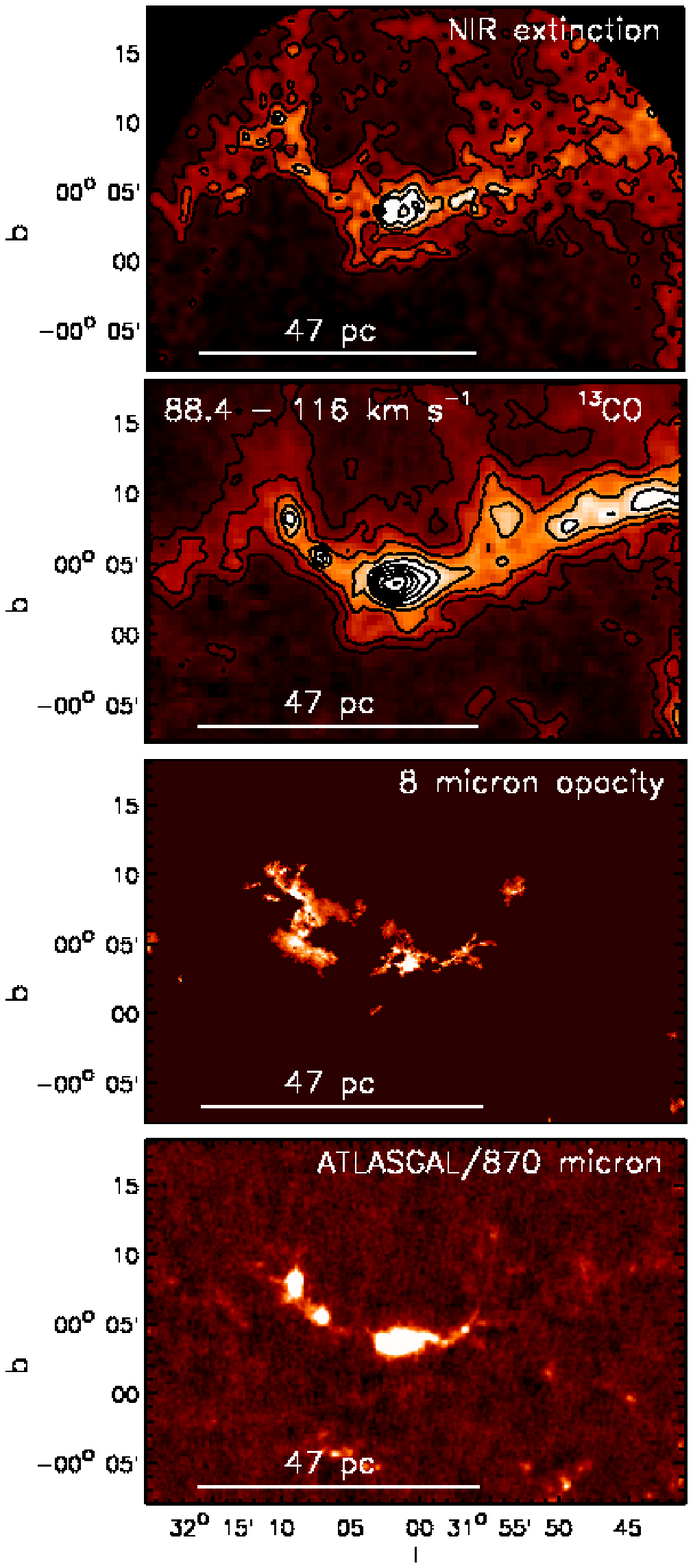}
      \caption{Same as Fig. \ref{fig:36.6}, but for IRDC complex 32.09+00.09. The 8 $\mu$m color scale saturates white at $A_\mathrm{V} \approx 20$ mag and the color scale of the 870 $\mu$m map at $A_\mathrm{V} \approx 7$ mag. 
              }
         \label{fig:32_0}
   \end{figure*}

   \begin{figure*}
   \centering
\includegraphics[bb = 155 20 500 700, clip=true, width=0.6\textwidth]{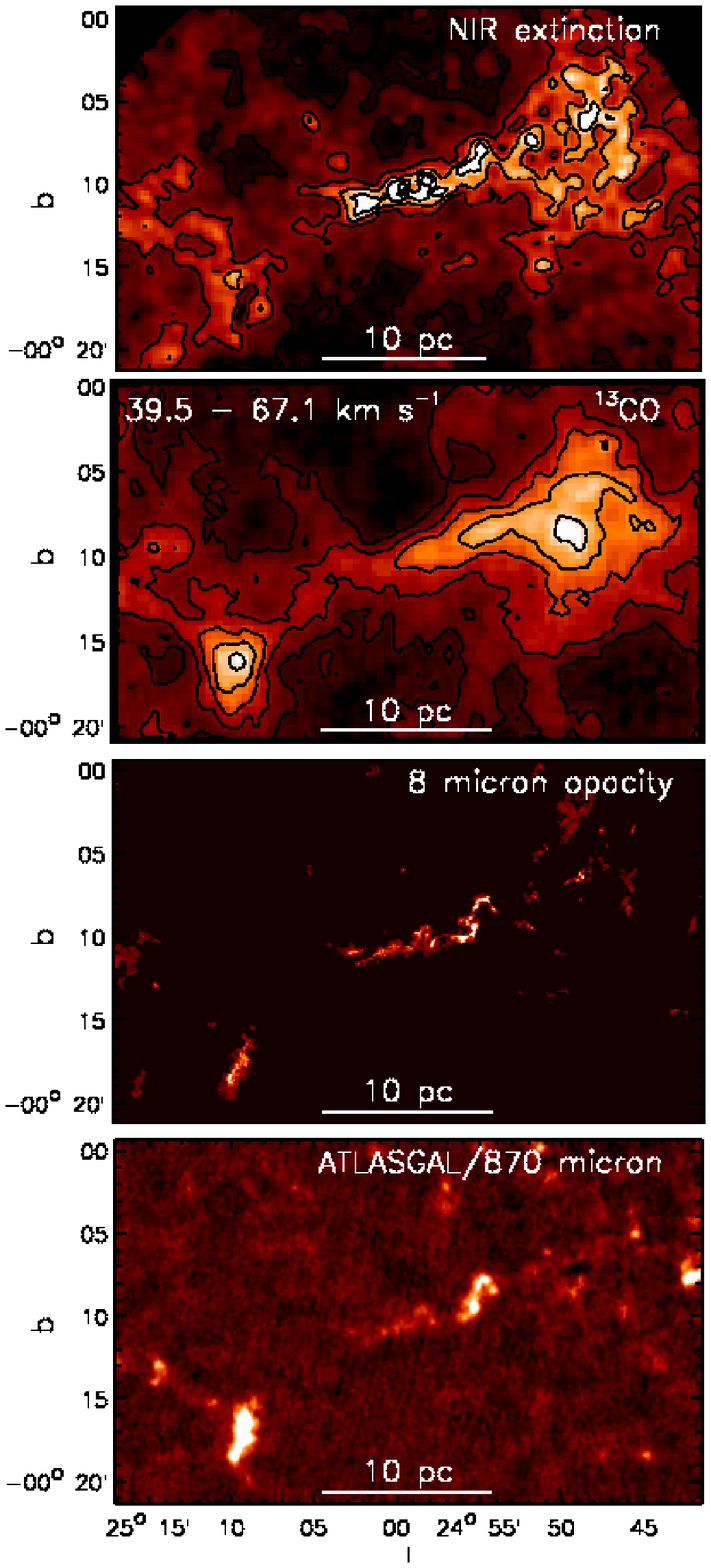}
      \caption{Same as Fig. \ref{fig:36.6}, but for IRDC complex 24.94-00.16. The 8 $\mu$m color scale saturates white at $A_\mathrm{V} \approx 40$ mag and the color scale of the 870 $\mu$m map at $A_\mathrm{V} \approx 7$ mag. 
              }
         \label{fig:25_0}
   \end{figure*}

   \begin{figure*}
   \centering
\includegraphics[bb = 80 20 570 700, clip=true, width=0.95\textwidth]{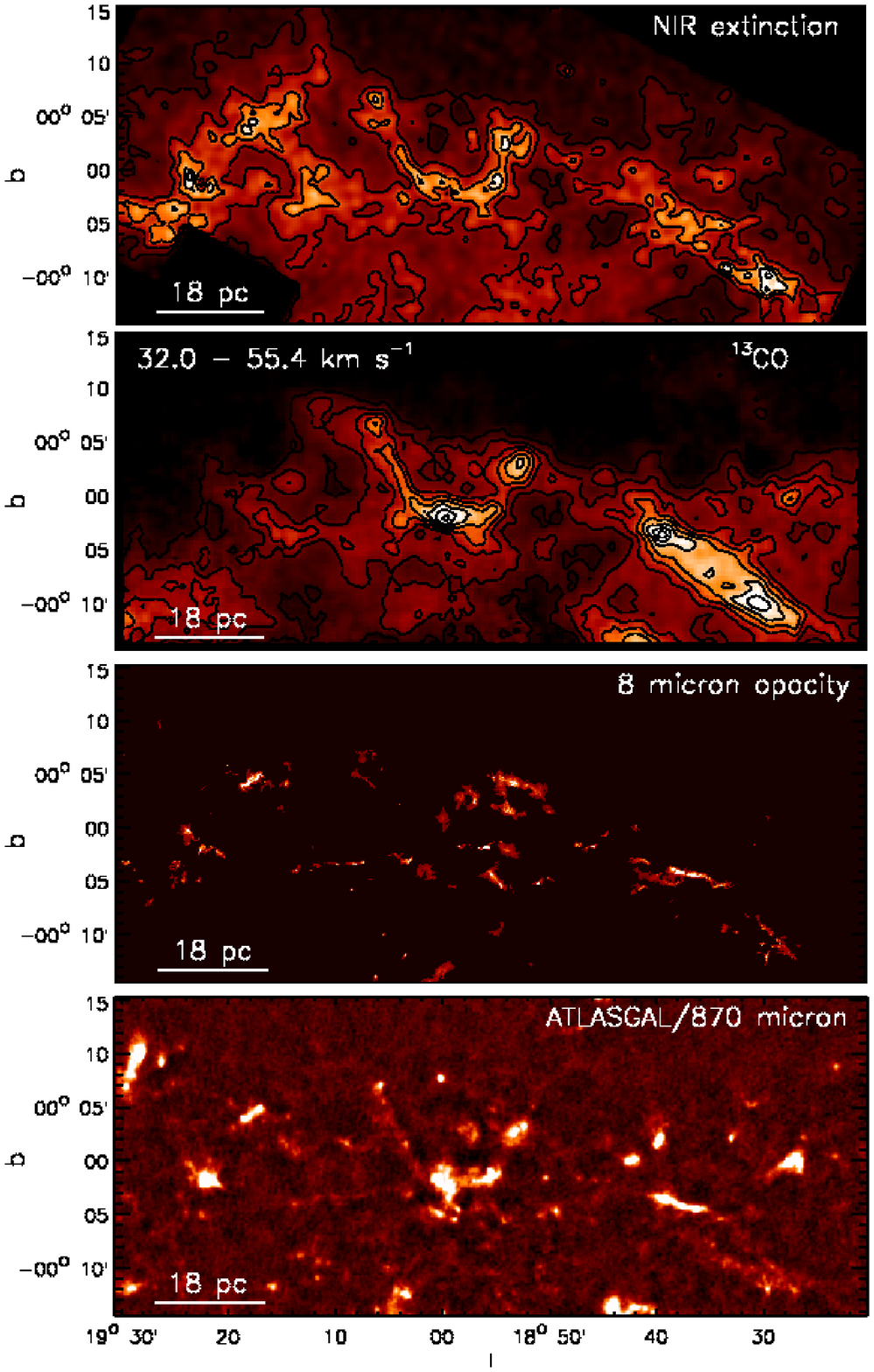} 
      \caption{Same as Fig. \ref{fig:36.6}, but for IRDC complex 18.54-00.16. The 8 $\mu$m color scale saturates white at $A_\mathrm{V} \approx 30$ mag and the color scale of the 870 $\mu$m map at $A_\mathrm{V} \approx 7$ mag. 
              }
         \label{fig:18_9}
   \end{figure*}

   \begin{figure*}
   \centering
\includegraphics[bb=90 50 580 700, clip=true, width=\textwidth]{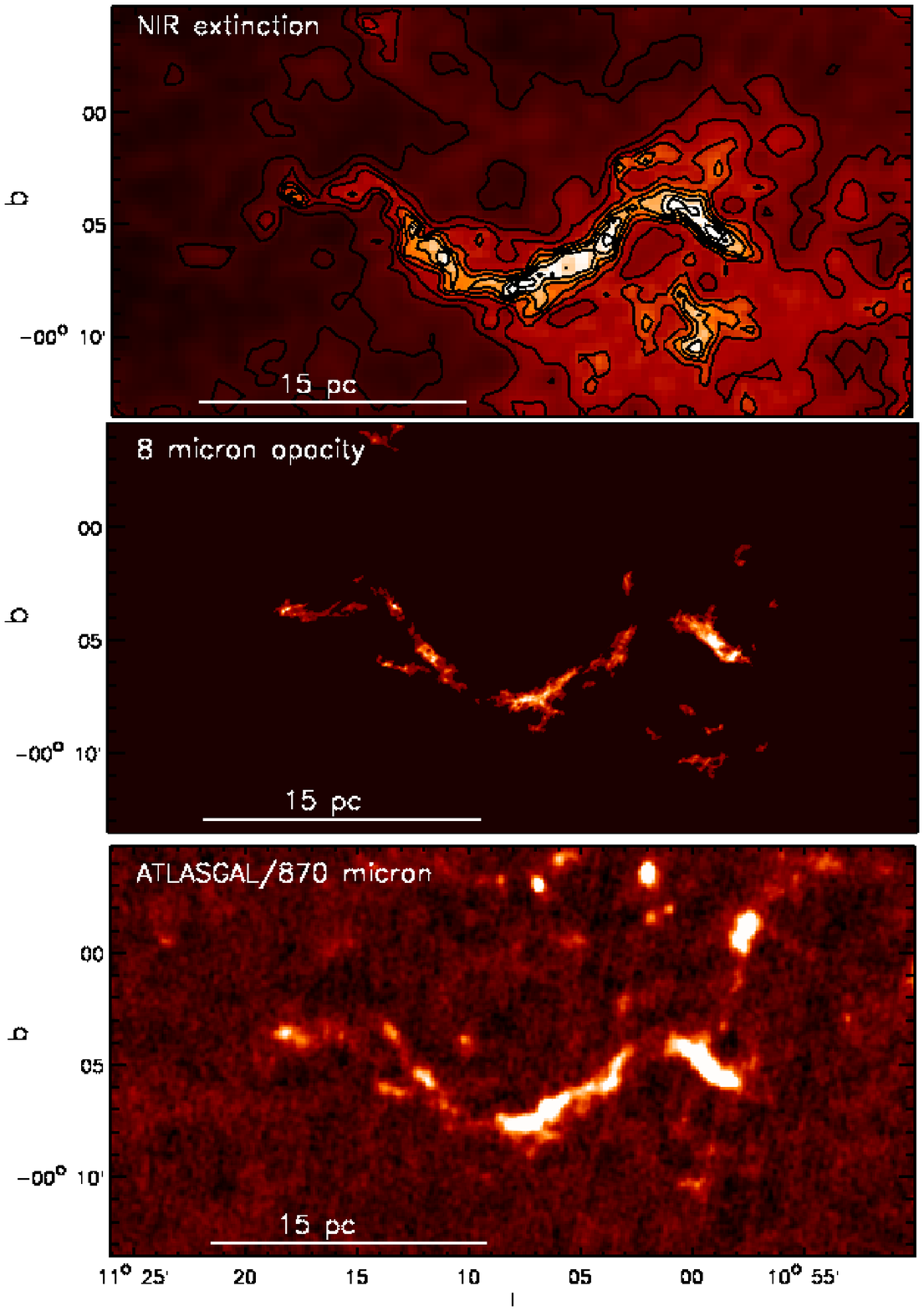}
      \caption{Same as Fig. \ref{fig:36.6}, but for IRDC complex 11.10-00.10. In addition to NIR extinction, only 8 $\mu$m dust opacity and ATLASGAL data are available for this cloud. The 8 $\mu$m color scale saturates white at $A_\mathrm{V} \approx 25$ mag and the color scale of the 870 $\mu$m map at $A_\mathrm{V} \approx 7$ mag. 
              }
         \label{fig:11_1}
   \end{figure*}

   \begin{figure*}
   \centering
\includegraphics[bb=80 30 580 680, clip=true, width=\textwidth]{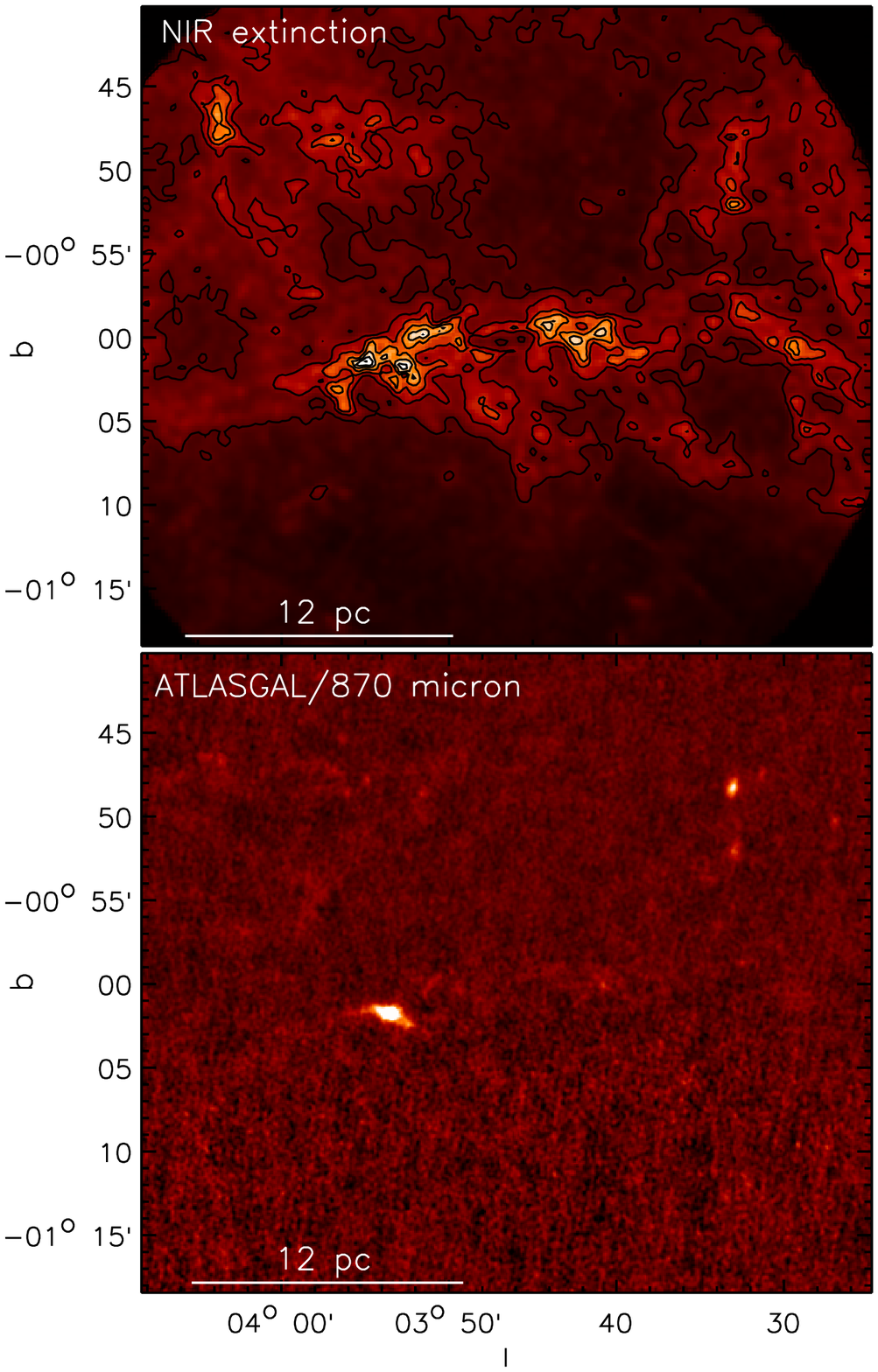}
\caption{Same as Fig. \ref{fig:36.6}, but for IRDC complex 03.80-01.00. In addition to NIR extinction, only ATLASGAL data are available for this cloud. The 8 $\mu$m color scale saturates white at $A_\mathrm{V} \approx 25$ mag and the color scale of the 870 $\mu$m map at $A_\mathrm{V} \approx 7$ mag. 
              }
         \label{fig:3_8}
   \end{figure*}

\end{document}